%% file: bare_jrnl.tex
\documentclass[journal]{IEEEtran}
\usepackage{comment}
\usepackage{amsmath}
\newcounter{MYtempeqncnt}
\usepackage{amsmath,amsfonts,amssymb,amsthm}
\usepackage{algorithmic}
\usepackage{algorithm}
\usepackage{multirow}
\usepackage{graphicx}
\usepackage{cite}
\usepackage{picinpar}
\usepackage{url}
\usepackage{flushend}
\usepackage{colortbl}
\usepackage{soul}
\usepackage{pifont}
\usepackage{color}
\usepackage{alltt}
\usepackage{enumerate}
\usepackage{breakurl}
\usepackage{pbox}
\usepackage{subfigure}
\usepackage{stfloats}
\usepackage{hyperref}
\hypersetup{
    colorlinks=true,
    linkcolor=blue,
    filecolor=magenta,      
    urlcolor=cyan,
}
 
\usepackage{pdfpages}
\pdfoutput=1
\usepackage{fancybox,framed}
\usepackage{csquotes}

\ifCLASSINFOpdf
\else
\fi
\begin{document}
\begin{titlepage}
\centering
\doublebox{%
\begin{minipage}{6in}
\begin{center}
This is the accepted version
\end{center}

\textbf{Live link to published version in IEEE Xplore:} \url{https://ieeexplore.ieee.org/document/8561163}
\newline
\newline
\textbf{Citation to the original IEEE publication:} A. Thakallapelli, S. Ghosh, and S. Kamalasadan, \enquote{Development and Applicability of Online Passivity Enforced Wide-Band Multi-Port Equivalents For Hybrid Transient Simulation}, IEEE Trans. on Power Syst., Volume: 34, No: 3, pp. 2302 {-} 2311, May. 2019.
\newline
\newline
\textbf{Digital Object Identifier (DOI): 10.1109/TPWRS.2018.2885240}
\newline
\newline
The following copyright notice is displayed here as per Operations Manual Section 8.1.9 on Electronic Information Dissemination (known familiarly as "author posting policy"):
\newline
\newline
\textcopyright{ 2018 IEEE}. Personal use of this material is permitted. Permission from IEEE must be obtained for all other uses, in any current or future media, including reprinting/republishing this material for advertising or promotional purposes, creating new collective works, for resale or redistribution to servers or lists, or reuse of any copyrighted component of this work in other works.
\end{minipage}}

\end{titlepage}
%
\title{Development and Applicability of Online Passivity Enforced Wide-Band Multi-Port Equivalents For Hybrid Transient Simulation}

%
%
%

\author{Abilash~Thakallapelli,~\IEEEmembership{Student~Member,~IEEE,} Sudipta~Ghosh,~\IEEEmembership{Member,~IEEE,} and~Sukumar~Kamalasadan,~\IEEEmembership{Senior~Member,~IEEE}
\thanks{A. Thakallapelli, S. Ghosh, and S. Kamalasadan (corresponding author), are with the Power, Energy and Intelligent Systems Laboratory, Energy Production Infrastructure Center (EPIC) and Department of Electrical Engineering, University of North Carolina at Charlotte, Charlotte, NC 28223 USA (e-mail: athakall@uncc.edu, sghosh9@uncc.edu, skamalas@uncc.edu).}

}

\maketitle
\vspace{-20mm}
\begin{abstract}
This paper presents a method for developing single and multi-port Frequency Dependent Network Equivalent (FDNE) based on a passivity enforced online recursive least squares (RLS) identification algorithm which identifies the input admittance matrix in $z$-domain. Further, with the proposed architecture, a real-time hybrid model of the reduced power system is developed that integrate Transient Stability Analysis (TSA) and FDNE. Main advantages of the proposed architecture are, it identifies the FDNE even with unknown network parameters in the frequency range of interest, and yet can be implemented directly due to discrete formulation while maintaining desired accuracy, stability and passivity conditions. The accuracy and characteristics of the proposed method are verified by implementing on two-area, IEEE 39 and 68 bus power system models.
\end{abstract}
\vspace{-4mm}

%
\IEEEpeerreviewmaketitle
\vspace{-4mm}
\section*{Nomenclature}
\addcontentsline{toc}{section}{Nomenclature}

\begin{IEEEdescription}[\IEEEusemathlabelsep\IEEEsetlabelwidth{EMT+FDNE+TSA(AGG)}]
\item[EMT Model] The study and external areas are modeled as EMT type (Original model).
\item[EMT+TSA] The external  area  is  modeled  as  TSA  type  equivalent with only network aggregation.
\item[EMT+TSA(AGG)] The external  area  is  modeled  as  TSA  type with both network and generator aggregation.
\item[EMT+FDNE] The external  area  is  modeled as FDNE type generated using the proposed algorithm.
\item[EMT+FDNE (VF)] The external  area  is  modeled as FDNE using the Vector Fitting (VF) from literature.
\item[EMT+FDNE+TSA] The external  area  is  modeled as a  combination  of  FDNE  and  TSA  type  with  only network  aggregation.
\item[EMT+FDNE+TSA (AGG)] The external  area  is  modeled as  a  combination  of  FDNE  and  TSA  type  with  both  network  and  generator  aggregation.
\end{IEEEdescription}

\section{Introduction}
%
%
%
%
\IEEEPARstart{R}{eal-time} EMT simulation requires detailed modeling of transmission systems to understand the effect of transients and harmonics arising due to varying operating conditions and disturbances in power grid. Effect of power electronic components associated with renewable energy sources on the power grid and performance of different controllers can be analyzed using EMT simulations \cite{ref1a}. Typically, integration time step of EMT simulation is in microseconds ($\mu s$). This makes modeling of a large transmission system for EMT studies impractical as detail modeling increases complexity and computational burden. One solution is to model the transmission system as TSA type with larger integration time as TSA simulations can run faster than EMT. However, in TSA type, due to large integration time, high-frequency transients following a disturbance in the system are not preserved making this approach not very accurate \cite{ref2a}. 

Another approach is to model large transmission network as frequency dependent reduced order systems that can represent the power grid under any operating condition. One way to reduce large power grid is to model part of the grid which is of interest (“study area”) in detail and the remainder of the system (“external area”) by an efficient equivalent such as FDNE \cite{ref6}. In this process, initially, the network is divided into study and external area based on the coherency grouping of generators such that all coherent generators are present in the external area \cite{ref19}. The boundary between the study and external area is divided considering the fact that interconnecting points should have the least minimum number of ports. Generally, in TSA type equivalent, the network admittance is evaluated only at the fundamental frequency, hence this representation ignores high-frequency oscillations. The high-frequency behavior of the external area can be preserved by using FDNE \cite{ref3a}, however, FDNE ignores electromechanical oscillations. In order to cover both electromechanical and high-frequency behavior, the external area should be modeled as a combination of TSA equivalent and FDNE.

FDNEs are generally formulated as frequency-dependent black-box terminal equivalents based on rational functions. In \cite{ref1}, FDNE is formulated using modal vector fitting (VF) method and in  \cite{ref2} passivity of the rational function is enforced using mode revealing algorithm. The authors in \cite{ref3}, used VF method to generate FDNE combined with TSA component for real-time simulation and, in \cite{ref4} this method was improved by including generator coherency. Further, the applicability of these models for large power system real-time simulation is studied in \cite{ref5}. Formulation of FDNE using time domain VF and simulated time domain response evaluations are performed in \cite{ref6}, and in \cite{ref7}, authors compared different rational approximation methods for simulated time domain responses. A rational approximation approach for formulating FDNE based on matrix pencil method is proposed in \cite{ref8} where as \cite{ref9} proposed a method for electromechanical and electromagnetic transient analysis using FDNE. 

However, state-of-the-art algorithms formulate FDNE with continuous domain transfer function which makes it very complex to implement these algorithms in the real-time simulator (EMT based).
 Also, in many of these methods FDNE formulation is dependent on the availability of admittance data over a wide range of frequency which is not always possible. Further, off-line calculation of admittance over a wide frequency range, and storing and retrieving of the data for curve fitting is tedious and time consuming. For example, if the frequency range of interest is from 0 to 5000 Hz with a step size of 0.1 Hz, one should construct 50001 admittance matrices at each frequency sample which is extremely complex to compute. This complexity increases with sample size, number of ports, and size of network under consideration.

This paper introduces a novel method for formulating FDNEs based on online RLS identification. In this method, the external area is energized with constant voltage source at the boundary buses after all the voltage sources and current sources in the network are short-circuited and open circuited respectively. Subsequently, by tracking input voltage and output current, FDNE is formulated. The proposed method simplifies FDNE formulation as it is independent of the availability of network parameters over a wide frequency range and its formulation is in discrete domain directly. Also the computational burden is reduced by using Kron's node elimination method in the external area network. The advantages of the proposed architecture are:
\begin{itemize}
    \item FDNE formulation is independent of the availability of network parameters over a wide frequency range.
    \item The architecture formulates FDNE in discrete domain, which reduces complexity in interfacing FDNE with the real-time simulator.
    \item The architecture formulates coherency based network equivalents of complex networks with less computational burden and desired accuracy.
    \item The methodology enforces stability and passivity conditions to ensure stable EMT simulations.
    \item The methodology can be directly implemented for real-time control \cite{ref10}.
\end{itemize}

The rest of the paper is organized as follows: In section II overall architecture is discussed. In section III implementation test on the interconnected power grid is discussed and Section IV concludes the paper. 
\vspace{-4mm}
\section{Proposed Methodology}
In large power grids, it has been frequently observed that, after disturbances, the generators swing together in groups, meaning units near a disturbance respond faster and together, whereas distant machines show relatively damped oscillatory behavior. This physical property is known as \textit{coherency} and group of machines with similar responses are termed coherent generators. In our work, the generators are coherently grouped based on a localness index.  Further,  power system model order reduction is performed by dividing the original system into a \textit{study} area and an \textit{external} area. The proposed method further divides the external area into two parts. The first one is a low-frequency equivalent (TSA) and the second one is high-frequency equivalent (FDNE). The reduced order modeling of power system involves the following steps:
\vspace{-3.5mm}

\subsection{Aggregation of External Area for TSA type modeling and Real-time Integration}
\begin{figure}[!h]
\centering
\includegraphics[width=3.5in,height=1.8in]{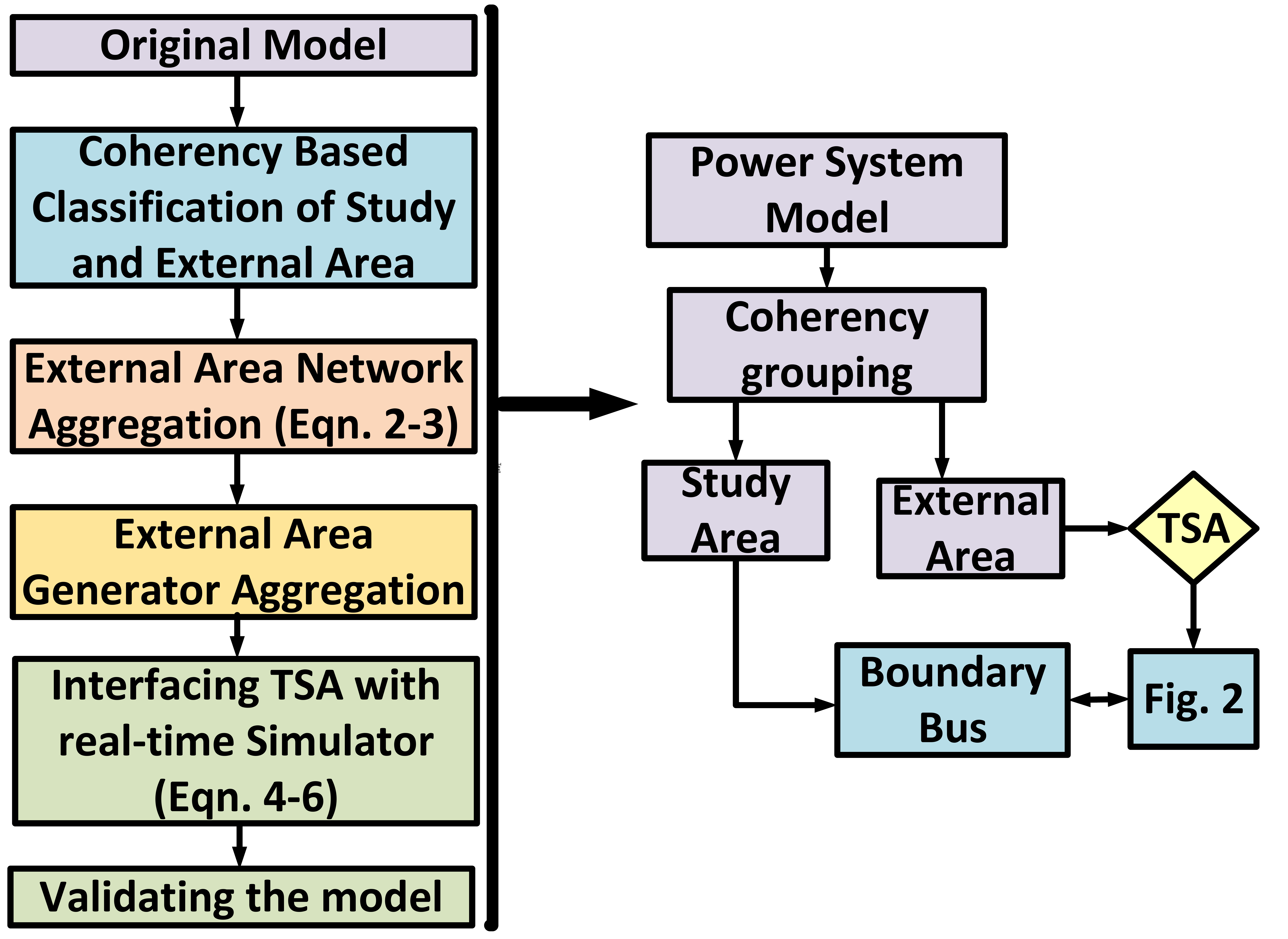}
\caption{The conceptual and functional flowchart for  TSA type modeling}
\label{fig2}
\vspace{-2mm}
\end{figure}

For retaining the electromechanical behavior of the system under consideration, aggregated TSA model is used. Fig. \ref{fig2} shows the flowchart for TSA type modeling.
\vspace{-3.5mm}

\subsubsection{Coherency Based Classification of Study and External Area}
For coherency grouping of the power system under study, initially the small signal stability study of the power grid model is performed and the generator electromechanical modes of oscillation are evaluated. Further, based on participation factors of all the generators, a localness index is calculated as follows:
\begin{equation}
\label{eqn1}
L_{index,i}=\sum_{k=1}^{n}(1-P_{ki})^n
\end{equation}
where $n$ is a number of synchronous generators connected in the system and, $P_{ki}$  is the normalized participation factor of the $k^{th}$ machine in the $i^{th}$ mode. For example, Table. \ref{table1} shows the coherency grouping of generators for IEEE 39 bus system based on the localness index. More details of the localness index are discussed in \cite{ref11}. 
\begin{table}[!h]
\renewcommand{\arraystretch}{1.3}
\centering
\caption{Coherency grouping of generators for IEEE 39 bus system}
\label{table1}
\begin{tabular}{*9c}
\hline
\textbf{Group} &  I &  II &  III &  IV\\
\hline
Generators & 4,5,6,7,9 & 1,10,8 & 3 & 2\\
\hline
\end{tabular}
\end{table}

\subsubsection{Network Aggregation}
For aggregating the external area network, the admittance matrix $(Y_{n\times n})$ of the external area is formulated using the bus and line data (at 60Hz). For example, if there are $n$ number of buses in the external area and we want to retain $m$ buses (i.e. $m=i+j$ , where $j$  be the number of boundary buses and $i$ is the number of generator buses) and eliminate the remaining $n-m$  buses in the external area, then using Kron node elimination method \cite{ref12}, reduced admittance matrix $(Y_{red})$ can be obtained as:
\begin{equation}
\label{eqn2}
Y_{red(m\times m)}=\left[Y_{m\times m}-Y_{m\times n}Y_{n\times n}^{-1}Y_{n\times m}\right]
\end{equation}
\begin{eqnarray}
   \left[\begin{array}{c}
    I_{b(j\times 1)} \\
    I_{g(i\times 1)}
   \end{array}\right]=
   Y_{red(m\times m)} \left[\begin{array}{c}
    V_{b(j\times 1)} \\
    V_{g(i\times 1)}
   \end{array}\right]
\label{eqn3}
\end{eqnarray}
where subscript $b$ and $g$ represents the boundary and generator buses respectively. 
\subsubsection{Generator and Associated Controller Aggregation}
After network aggregation and generators are left intact, the reduced admittance matrix is of the size $m\times m$ ($m=i+j$, i.e.  there are $j$ boundary buses and $i$ generator buses). To further reduce computational burden and to reduce complexity in modeling, generators and associated controllers can be aggregated. With generator aggregation, the reduced admittance matrix is of the size $(m-i+1)\times (m-i+1)$  ($m=i+j$, i.e. $i$ coherent generators can be aggregated into one generator). In the case where generators are not aggregated the accuracy of coherency grouping has no significant effect on the TSA, but in the case of generator aggregation, all the generators which are to be aggregated must be in the same coherent group. For scenarios where multiple operating condition changes, the grouping can be performed using online coherency grouping \cite{ref4a}.
Method of generator and controller aggregation is discussed in \cite{ref13}. Thus additional details are not explained in the paper.

\subsubsection{Interfacing TSA type modeling with real-time simulator}
\begin{figure}[h!]
\centering
\includegraphics[width=3.5in,height=2.5in]{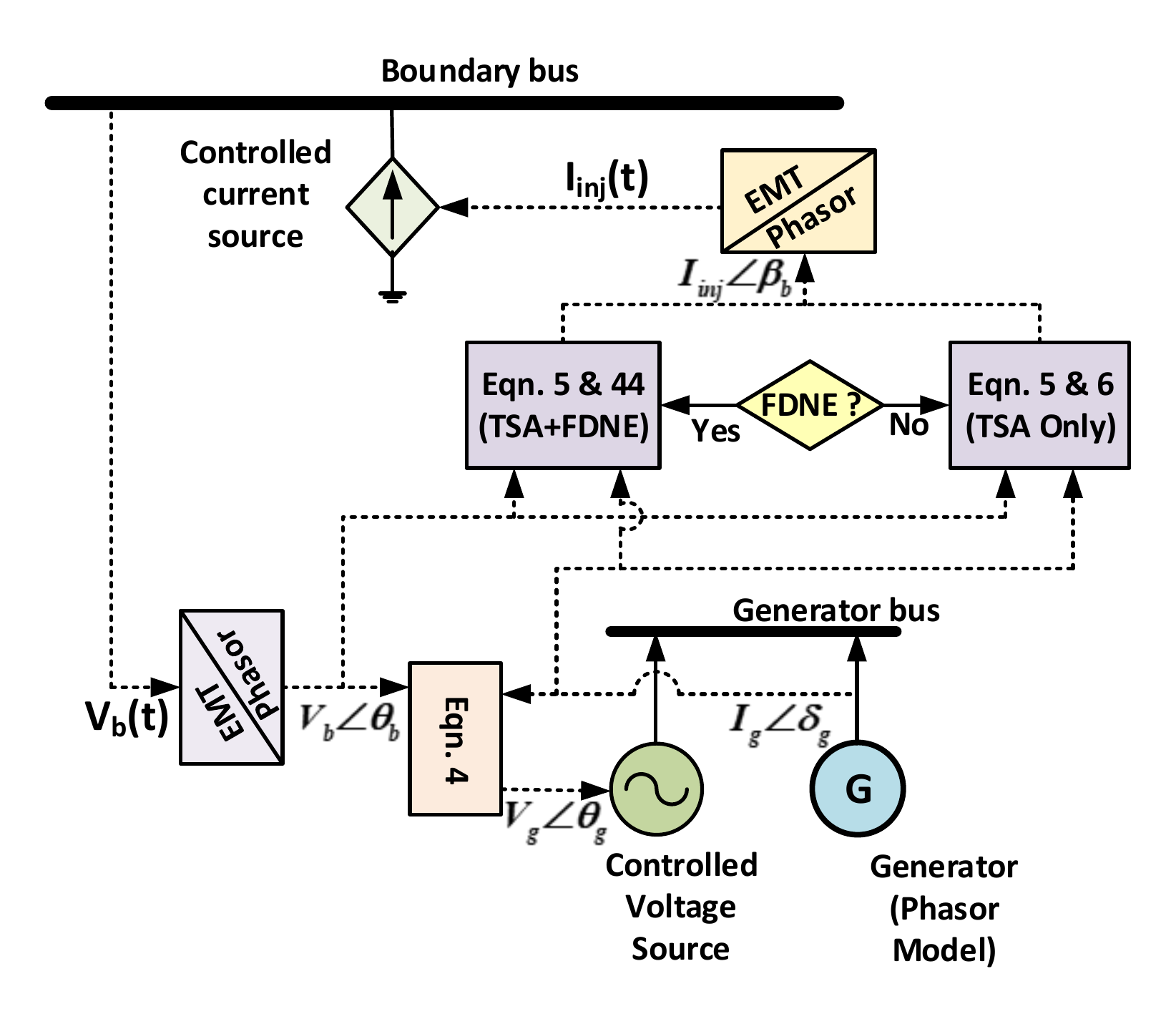}
\vspace{-10mm}
\caption{TSA Calculation}
\label{fig2x}
\end{figure}

In this step, voltages at the boundary buses are the input to the TSA block, where output currents from the TSA block are injected back to the boundary buses. Here, the generators are modeled in detail to observe the electromechanical behavior. Conversion of boundary bus voltage from time domain to phasor domain is then performed using discrete sequence analyzer \cite{ref14}. The result gives magnitude $|V_b|$  and, phase angle $\angle V_b$. Also, phase angle of $V_b$  with reference to $I_b$ can be determined as $\theta_b=\angle \delta_b-\angle V_b$, where $\angle \delta_b$   is the angle of $I_b$. Then, using \eqref{eqn4} the generator bus voltage can be calculated as
\begin{equation}
\label{eqn4}
V_g\angle \theta_g = \left(I_g\angle \delta_g-Y_{gb}V_b\angle \theta_b\right)Y_{gg}^{-1}
\end{equation}
where $I_g$ is the generator current injection, $\angle \delta_g$  is the angle of generator current, $V_b$  is the boundary bus voltage, $V_g$  is the generator bus voltage and, $\angle \theta_g$ is the angle of generator bus voltage. Using $I_g$ (for the first iteration the initial value is obtained from the power flow solution) and $V_b$, the generator voltage $V_g$  is calculated. The generator bus is energized with $V_g$ and subsequently $I_g$ is obtained from generator phasor model. From calculated $V_g$  and boundary bus voltage $V_b$, boundary bus current injection $I_{inj}$  is calculated as shown in \eqref{eqn5}-\eqref{eqn6}. A graphical description is shown in Fig. \ref{fig2x}.
\vspace{-3mm}
\begin{equation}
\label{eqn5}
I_b\angle \delta_b = Y_{bb}V_b\angle \theta_b+Y_{bg}V_g\angle \theta_g
\end{equation}
\begin{equation}
\label{eqn6}
I_{inj}\angle \beta_b = I_b\angle \delta_b
\end{equation}
After calculating boundary bus current $I_{inj}\angle\beta_b$  in phasor form, it is converted into the time domain and injected into the boundary bus. The overall implementation is shown in Fig. \ref{fig2}.

\vspace{-4mm}

\subsection{Study and analysis of TSA type equivalent modeling}\label{sec11}

\begin{figure}[h]
\centering
\includegraphics[width=3.5in,height=1.5in]{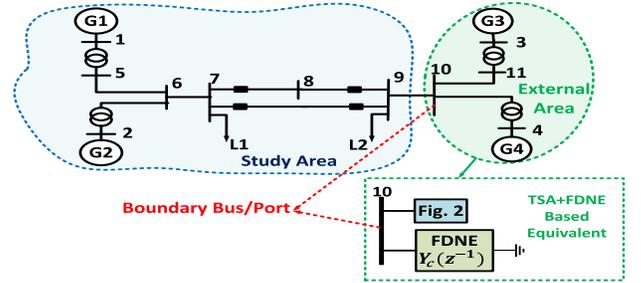}
\caption{Proposed dynamic equivalent of two area system}
\label{fig4}
\vspace{-3mm}
\end{figure}

For study and preliminary analysis, the TSA type equivalent is implemented on two area power system model as shown in Fig. \ref{fig4}. In this test system, area-1 consists of generators G1, G2, and area-2 consists of generators G3, G4 \cite{ref15}. For analysis purpose, area-2 is considered as external area with boundary bus as bus 10. Two cases are considered here. In the first case \textit{EMT+TSA Based Model} and in the second case \textit{EMT+TSA Based Model (AGG)} are analyzed. For bench-marking, both the test cases are compared with (\textit{EMT Based Model}). 

\begin{figure}[h]
\centering
\includegraphics[width=3.5in,height=1.4in]{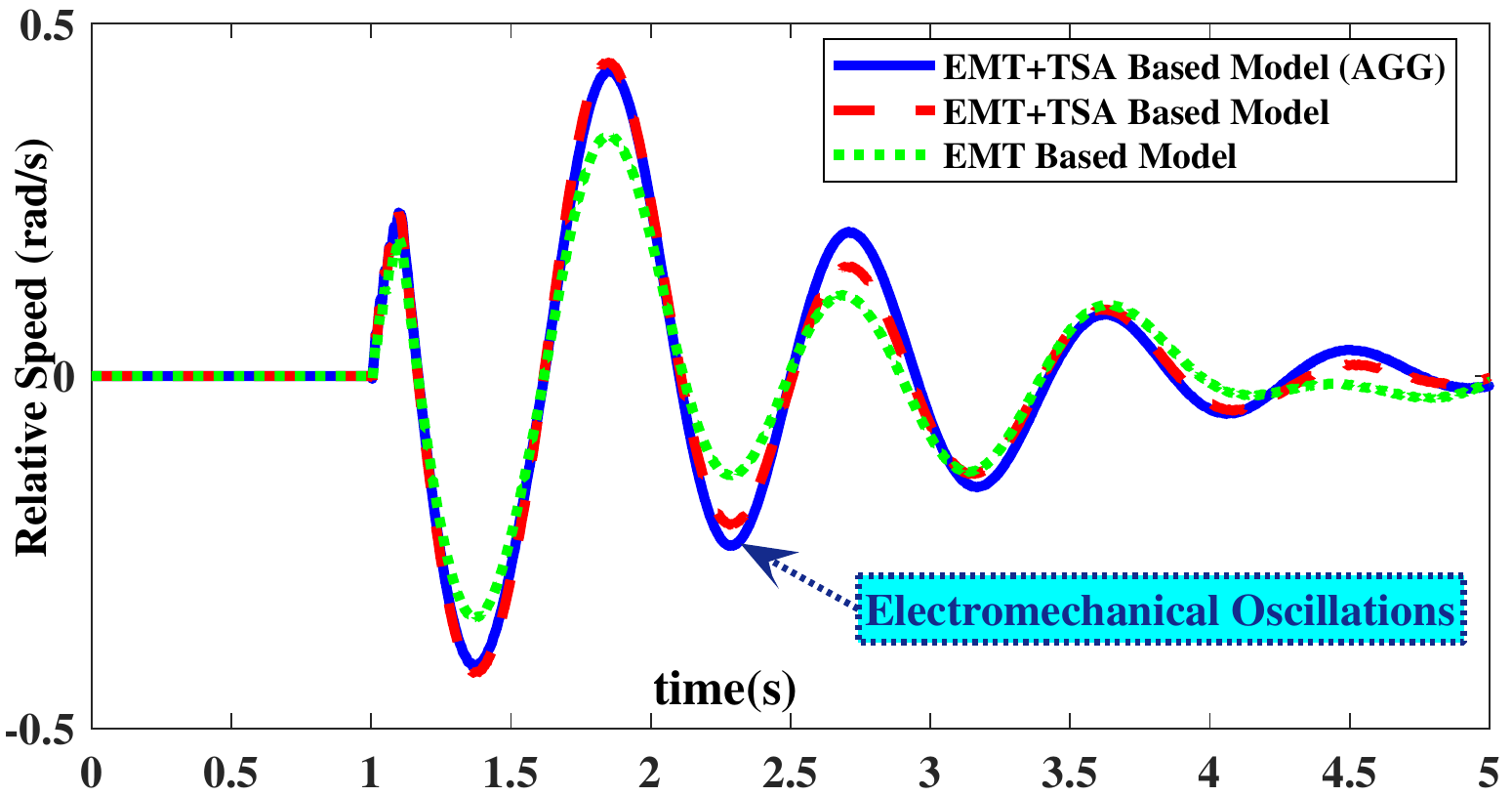}
\caption{The relative speed of Gen.2 w.r.t Gen.1}
\label{fig5}
\vspace{-4mm}
\end{figure}

\begin{figure}[h]
\centering
\includegraphics[width=3.5in,height=1.4in]{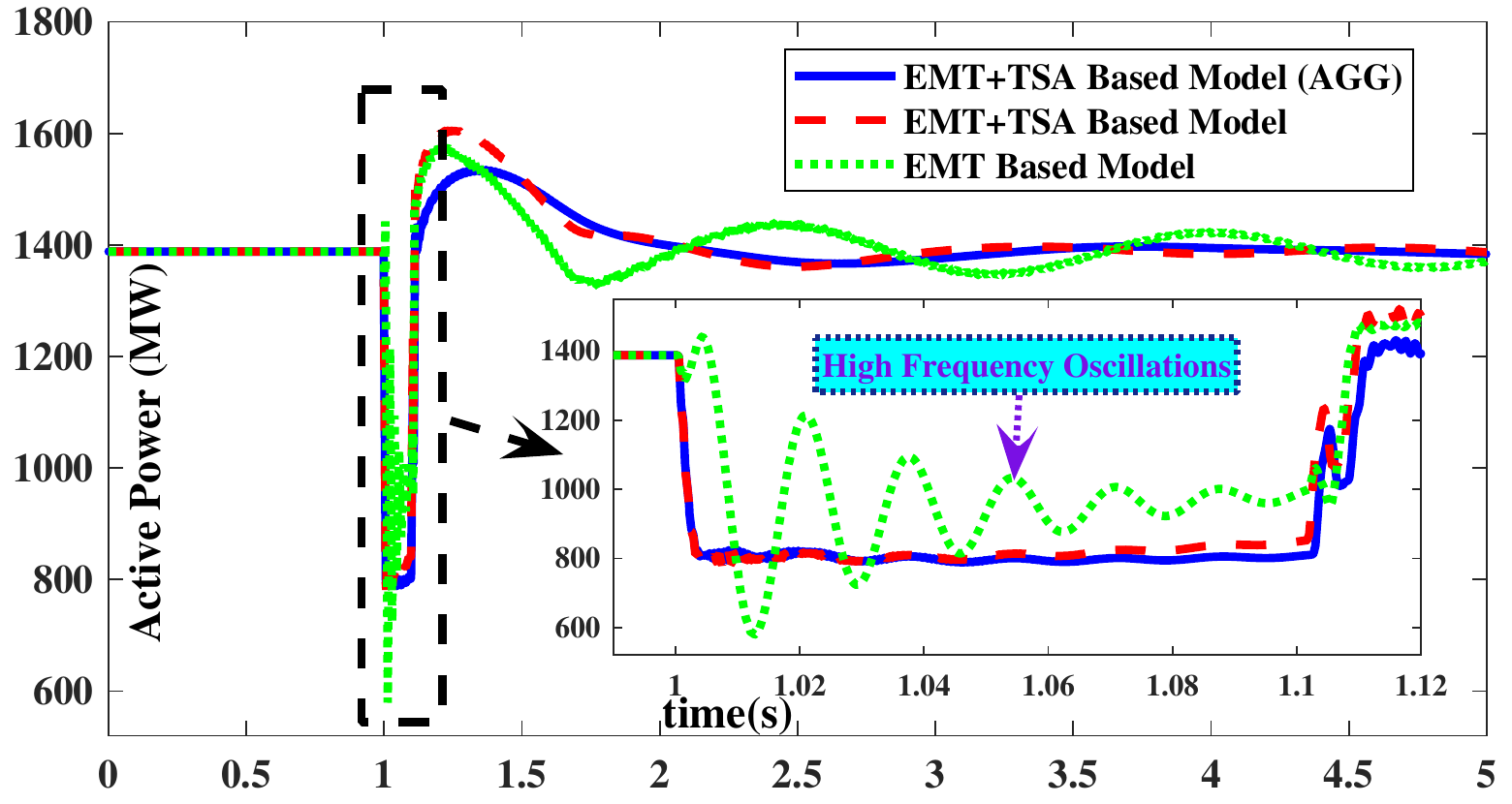}
\caption{Active power flow from bus 10 to bus 9}
\label{fig6}
\vspace{-4mm}
\end{figure}

For validating the proposed approach a 3-ph fault is created on Bus-8 at 1s for a duration of 0.1s. Fig. \ref{fig5} shows the comparison of the relative speed of Gen. 2 with respect to (w.r.t) gen. 1, and Fig. \ref{fig6} shows the comparison of active power flow from bus 10 to bus 9. In general, after a disturbance, high-frequency transients occur for a short duration whereas electro-mechanical oscillation will be for longer duration. From Fig. \ref{fig5}, it can be observed that due to large simulation time step compared to EMT type, TSA type equivalent can only preserve electro-mechanical oscillations, but high-frequency oscillations are not preserved as it can be seen from Fig. \ref{fig6}.

For quantitative analysis, relative error between the two cases and full EMT type model is calculated using \eqref{eqn44}. The results are tabulated in Table. \ref{table1aa}.
\begin{equation}
\label{eqn44}
relative\ error = \frac{\left\|y_{ref}-y_{act}\right\|_2}{\left\|y_{ref}\right\|_2}
\end{equation}
where $y_{ref}$ represents the output from full EMT model (\textit{EMT Based Model}) and $y_{act}$  represents the output obtained in each case (\textit{EMT+TSA Based Model (AGG)} and (\textit{EMT+TSA Based Model}) respectively)
\vspace{-5mm}
\begin{table}[!h]
\renewcommand{\arraystretch}{1.3}
\centering
\caption{Comparison of Reduced (EMT+TSA) and Original (EMT) Models}
\label{table1aa}
\begin{tabular}{*9c}
\hline
 & EMT+TSA(AGG) & EMT+TSA \\
\hline
Relative Speed (Fig. \ref{fig5}) & 0.3589 & 0.2998 \\
\hline
Active Power (Fig. \ref{fig6}) & 0.0356 & 0.0351 \\
\hline
\end{tabular}
\end{table}
\vspace{-3mm}

It can be seen that the aggregated model has a large error. This motivates the use of FDNE representations.
\vspace{-3mm}
\subsection{FDNE Formulation and Real-time simulator Integration}
\begin{figure}[!t]
\centering
\includegraphics[width=3.5in,height=2.2in]{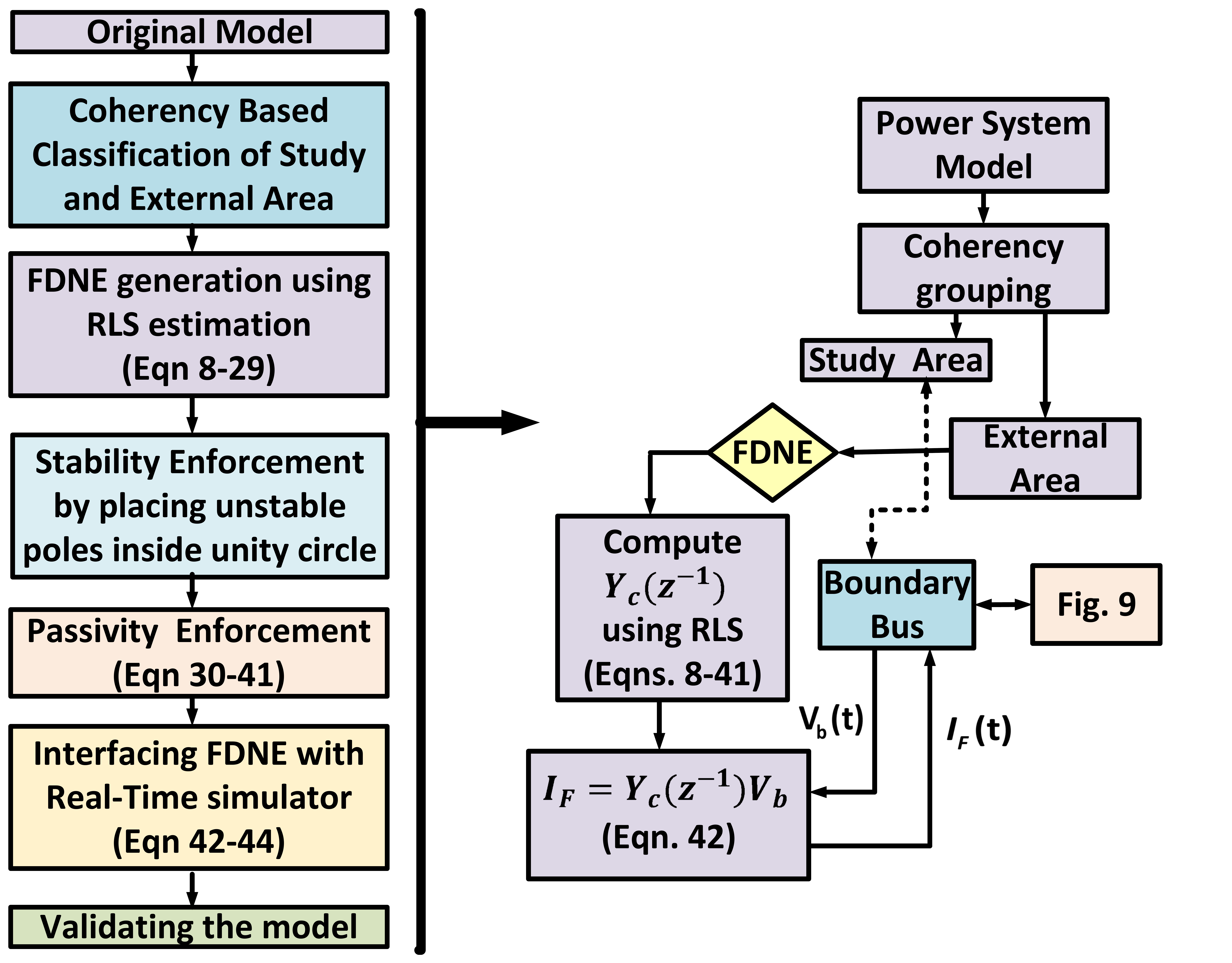}
\caption{The conceptual and functional flowchart for  FDNE modeling}
\label{fig7}
\vspace{-7mm}
\end{figure}

\vspace{-1mm}
For retaining the high-frequency behavior of the system under consideration, FDNE is formulated. Fig. \ref{fig7} shows the FDNE modeling flowchart. In this method, the external area is energized with a constant voltage source with varying frequency in steps of 0.01 (from few Hz to kHz) after short-circuiting all voltage sources, and open circuiting all current sources. Since, the goal here is to identify the frequency dependent admittance $(Y_{f})$, the boundary bus voltage and current are the required signals. FDNE is then formulated in $z$-domain using RLS \cite{ref16,ref17} by tracking input voltage and output current. The basic principle is as follows. If $V_F$  is the voltage input to the boundary bus and $I_F$  is the current output from the boundary bus, then $Y_{f}$ can be written as
\begin{equation}
\label{eqn7}
Y_{f}(z^{-1})=\frac{I_F(k)}{V_F(k)}=\frac{b_1z^{-1}+b_2z^{-2}+...+b_nz^{-n}}{1+a_1z^{-1}+a_2z^{-2}+...+a_nz^{-n}}
\end{equation}
where $k$  is the number of samples.
For a $m$-port network (that means $m$ boundary buses), $Y_{f}$  can be represented as in \eqref{eqn8}, where $Y_{f(m,m)}$  and $Y_{f(m,p)}$ in \eqref{eqn8} are the self and mutual admittance respectively. FDNE model formulation and validation involves the following steps.
\begin{figure*}[!h]
\normalsize
\setcounter{MYtempeqncnt}{\value{equation}}
\begin{equation}
\label{eqn8}
Y_{f}(z^{-1})_{m\times m}=\left[\begin{array}{cccc}
    Y_{f(1,p)}+Y_{f(1,1)}+\dots+Y_{f(1,m)} & -Y_{f(1,2)} & \dots & -Y_{f(1,m)} \\
    Y_{f(2,1)} & . & \dots & -Y_{f(2,m)} \\
    . & . & \dots & . \\
    -Y_{f(m,1)} & . & \dots & Y_{f(m,p)}+Y_{f(m,1)}+\dots+Y_{f(m,m)} \\
   \end{array}\right]
\end{equation}
\hrulefill
\end{figure*}

\subsubsection{Recursive Least Square Estimation}
Identification of a dynamic process is performed using the process input $u(k)$  and the process output $y(k)$  at every sample $k$ . Considering the $z$-domain model of an $n^{th}$  order process, this can be represented as
\begin{equation}
\label{eqn9}
\frac{y(k)}{u(k)}=\frac{b_1z^{-1}+b_2z^{-2}+\dots+b_nz^{-n}}{1+a_1z^{-1}+a_2z^{-2}+\dots+a_nz^{-n}}
\end{equation}
where $a's$ and $b's$ are the transfer function denominator and numerator coefficients respectively. 
For $N$  observation window length, \eqref{eqn9} can be rewritten as
\begin{eqnarray}
   \left[\begin{array}{c}
    y(k) \\
    y(k-1) \\
    . \\
    . \\
    . \\
     y(k-N+1)
   \end{array}\right]_{N\times1}=\left[X_{N\times2n}\right]
   \left[\begin{array}{c}
    a_1 \\
    . \\
    . \\
    a_n \\
    b_1 \\
    . \\
    . \\
    b_n
   \end{array}\right]_{2n\times1}
\label{eqn10}
\end{eqnarray}
Equation \eqref{eqn10} can be written in the generic form as
\begin{equation}
\label{eqn11}
\Phi_{model(N\times1)}=X_{N\times2n}\Theta_{2n\times1}
\end{equation}
where $X$ is a matrix of past inputs and outputs, $\Phi$ is a matrix of past and present outputs, and $\Theta$ is the coefficient matrix of the transfer function.

Assume that the model identified is different from measurements, then
\begin{equation}
\label{eqn12}
\epsilon=\Phi_{measured}-\Phi_{model}
\end{equation}
where $\epsilon$  is the error between the performance of the system measurement (subscript measured) and the model (subscript model). For reducing this error, a criteria $J$  can be defined as  
\begin{equation}
\label{eqn13}
J=\epsilon^t\epsilon
\end{equation}
By letting $dJ/d\Theta=0$,  we get
\begin{equation}
\label{eqn14}
\Theta=\left[X^tX\right]^{-1}X^t\Phi_{measured}
\end{equation}
From \eqref{eqn14} it can be seen that, in order to calculate the measured variable the inverse of the state matrix should be determined.  This can drastically slow down the process and some time may not be achievable. To circumvent this issue, an RLS based computational algorithm that eliminates the matrix inversion is designed.  Let $S=X^tX$, then \eqref{eqn14} can be written as 
\begin{equation}
\label{eqn15}
\Theta=S^{-1}X^t\Phi
\end{equation}
where $\Phi=\Phi_{measured}$
Then, 
\begin{eqnarray}
   \Theta(k)=S^{-1}\left[X(k)X^t(k-1)\right]
    \left[\begin{array}{c}
    \Phi(k) \\
    \Phi(k-1) 
   \end{array}\right]
\label{eqn16}
\end{eqnarray}
\begin{equation}
\label{eqn17}
\Theta(k)=S^{-1}\left[X(k)\Phi(k)+X^t(k-1)\Phi(k-1)\right]
\end{equation}
Using \eqref{eqn11}, \eqref{eqn17} can be written as
\begin{equation}
\label{eqn18}
\begin{aligned}
\Theta(k)=S^{-1}\left[X(k)\Phi(k)+X^t(k-1)X(k-1)\Theta(k-1)\right]
\end{aligned}
\end{equation}
\begin{equation}
\label{eqn19}
\Theta(k)=S^{-1}\left[X(k)\Phi(k)+S(k-1)\Theta(k-1)\right]
\end{equation}
\begin{equation}
\label{eqn20}
S(k)=S(k-1)+X(k)X^t(k)
\end{equation}
Substituting \eqref{eqn20} in \eqref{eqn19}
\begin{equation}
\label{eqn21}
\Theta(k)=S^{-1}\left[X(k)\Phi(k)+\{S(k)-X(k)X^t(k)\}\Theta(k-1)\right]
\end{equation}
\begin{equation}
\label{eqn22}
\begin{aligned}
\Theta(k)=\Theta(k-1)+[S(k-1)+X(k)X^t(k)]^{-1}X(k)\\
[\Phi(k)-X^t(k)\Theta(k-1)]
\end{aligned}
\end{equation}
Let $P(k)=S^{-1}(k)$. Then by using matrix inversion lemma, $P(k)$ can be represented as 
\begin{equation}
\begin{aligned}
\label{eqn23}
P(k)=P(k-1)\left[I-\frac{X(k)X^t(k)}{1+X^t(k)P(k-1)X(k)}\right]P(k-1)
\end{aligned}
\end{equation}
Let 
\begin{equation}
\begin{aligned}
\label{eqn24}
K(k)=\frac{X(k)}{1+X^t(k)P(k-1)X(k)}
\end{aligned}
\end{equation}
Then, $P(k)$ can be re-written as
\begin{equation}
\begin{aligned}
\label{eqn25}
P(k)=\left[I-K(k)X^t(k)\right]P(k-1)
\end{aligned}
\end{equation}
Substituting \eqref{eqn25} in \eqref{eqn22}, \eqref{eqn22} can be represented as
\begin{equation}
\begin{aligned}
\label{eqn26}
\Theta(k)=\Theta(k-1)+K(k)\left[\Phi(k)-X^t(k)\Theta(k-1)\right]
\end{aligned}
\end{equation}
With weighted least square, \eqref{eqn24} and \eqref{eqn25} can be presented as
\begin{equation}
\begin{aligned}
\label{eqn27}
K(k)=\frac{P(k-1)X(k)}{\gamma+X^t(k)P(k-1)X(k)}
\end{aligned}
\end{equation}
\begin{equation}
\begin{aligned}
\label{eqn28}
P(k)=\frac{\left[I-K(k)X^t(k)\right]P(k-1)}{\gamma}
\end{aligned}
\end{equation}
where $\gamma$ is the weighting factor.

\begin{figure}[!h]
\centering     
\subfigure[Magnitude of admittance]{\label{fig8}\includegraphics[width=1.72in,height=1.4in]{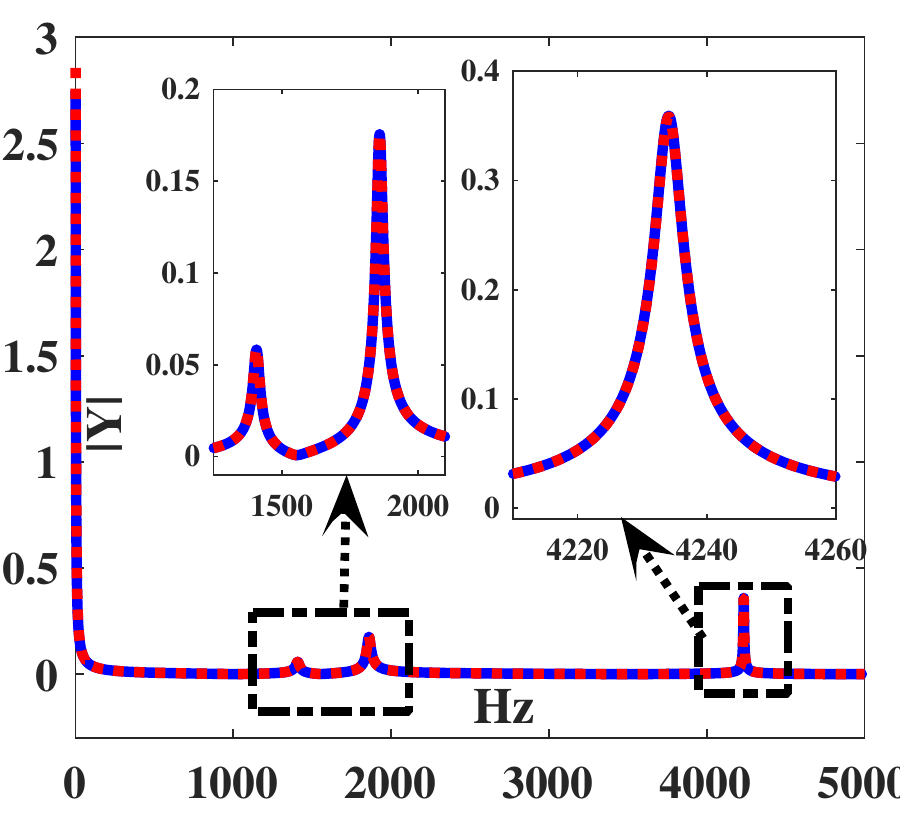}}
\vspace{-3.5mm}
\subfigure[Angle of admittance]{\label{fig9}\includegraphics[width=1.72in,height=1.4in]{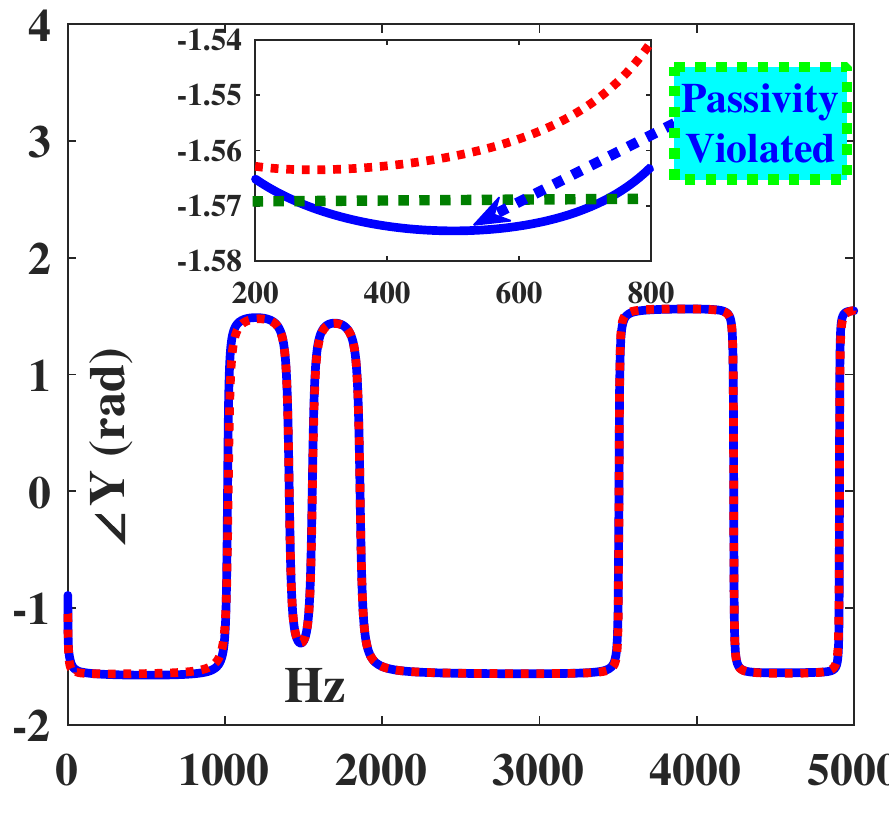}}
\vspace{-4mm}
\subfigure{\includegraphics[width=1.8in,height=0.15in]{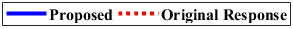}}
\caption{Admittance vs frequency of external area (two area system)}
\vspace{-4mm}
\end{figure}

Thus, with a given process input $V_F(k)$  and process output $I_F(k)$, $Y_{f}$  can be computed using RLS estimation. The validity of the proposed algorithm is verified by implementing on different test systems. In the first case, the proposed algorithm is implemented on two area test system with 1-port network. Fig. \ref{fig8} and Fig. \ref{fig9} shows the magnitude and angle of admittance of the external area and Table. \ref{table2} shows the comparison of FDNE formulation for two area system. It can be observed that even though this approach uses lower order transfer function (meaning less computational burden), it gives similar error compared to higher order VF method.

\begin{table}[!h]
\renewcommand{\arraystretch}{1.3}
\centering
\caption{FDNE Formulation Comparison}
\label{table2}
\begin{tabular}{*9c}
\hline
Type &  Proposed Method &  Vector Fitting(VF)\\
\hline
FDNE Order & 17 & 21\\
\hline
RMS Error & 2.176e-6 & 2.0101e-6\\
\hline
\end{tabular}
\vspace{-2mm}
\end{table}

\subsubsection{Passivity Enforcement}
For a stable EMT simulation, the admittance matrix should be passive. Presence of negative resistance (i.e angle of admittance $>$ $\pm 1.57\ rad$ in phasor form) in the frequency range of interest due to inherent approximations in identification violates passivity. For instance, from Fig. \ref{fig9} it can be seen that for the previous test case, passivity is violated. For enforcing passivity, an algorithm is developed as shown in Algorithm \ref{1a}. The details are as follows.

\begin{algorithm}
\caption{Algorithm for Passivity Enforcement}
\begin{algorithmic}[1] 
\STATE Calculate $Y_{f}(z^{-1})$  using RLS
\STATE $z=e^{j2\pi fT_s}$ where $f=1:f_{max}$, $f_{max}$ is the maximum frequency of interest, and $T_s$ is the sampling time
\STATE Obtain admittance matrix $(Y_{d})$ by substituting step-2 in step-1
\STATE Calculate $G_{d}$=$\Re\left[Y_{d}\right]$ from step-3
\STATE Calculate $G_{f}(z^{-1})$ and $B_{f}(z^{-1})$ $\left(eqns.\  \ref{eqn32}-\ref{eqn34}\right)$
\STATE Assume initial $G_{c}(z^{-1})=G_{f}(z^{-1})$
\FOR{$k=1$ to $length(f)$}
\IF{minimum eigenvalue of $G_{d}(k)$ $<0$}
\STATE using SDP calculate $G_{b}$
\STATE $\Delta G(k)=G_{b}(k)-G_{d}(k)$
\ELSE
\STATE $\Delta G(k) = 0$
\ENDIF
\STATE $G_{c}(z^{-1})=G_{c}(z^{-1})+\Delta G(k)$
\ENDFOR
\STATE Finally $Y_{c}(z^{-1})=G_{c}(z^{-1})+B_{f}(z^{-1})$ is obtained
\end{algorithmic}
\label{1a}
\end{algorithm}

From \eqref{eqn8}, let $Y_{f}(z^{-1})$ is the fitted admittance transfer function matrix which of the size $m\times m$ (where $m$  is the number of ports). By substituting $z=e^{i2\pi fT_s}$  in \eqref{eqn8}, the fitted admittance matrix data for $k$  frequency samples can be represented as
\begin{equation}
\label{eqn29}
\left[Y_{d}\right]_{m\times m\times k}=\left[Y_{f}(z^{-1})\right]_{m\times m}
\end{equation}
where $f\in1:f_{max}$ ($f_{max}$  being the maximum frequency under consideration) and, $T_s$  is the sampling time.
Let $G_{d}$  is the real part of the admittance matrix $(Y_{d})$. Then
\begin{equation}
\label{eqn30}
\left[G_{d}\right]_{m\times m\times k}=\Re \left[Y_{d}\right]_{m\times m\times k}
\end{equation}
For a function to be passive
\begin{equation}
\label{eqn31}
eig(G_{d})>0
\end{equation}
This implies that, if the admittance transfer function matrix is positive definite then it is also passive. Considering this, if the fitted function $Y_{f}$  violates \eqref{eqn31}, then a new corrected transfer function matrix $Y_{c}$  is formulated. The conductance transfer function $(G_{f})$ and susceptance transfer function ($B_{f}$) are calculated as follows.
\begin{equation}
\label{eqn32}
Y_{f}(z^{-1})=G_{f}(z^{-1})+B_{f}(z^{-1})
\end{equation}
\begin{equation}
\label{eqn33}
G_{f}(z^{-1})=\frac{1}{2}\left[Y_{f}(z^{-1})+Y_{f}(z^{-1})^*\right]
\end{equation}
\begin{equation}
\label{eqn34}
B_{f}(z^{-1})=\frac{1}{2}\left[Y_{f}(z^{-1})-Y_{f}(z^{-1})^*\right]
\end{equation}
where $*$ stands for complex conjugate.

\begin{figure}[!h]
\centering
\includegraphics[width=3.5in,height=1.5in]{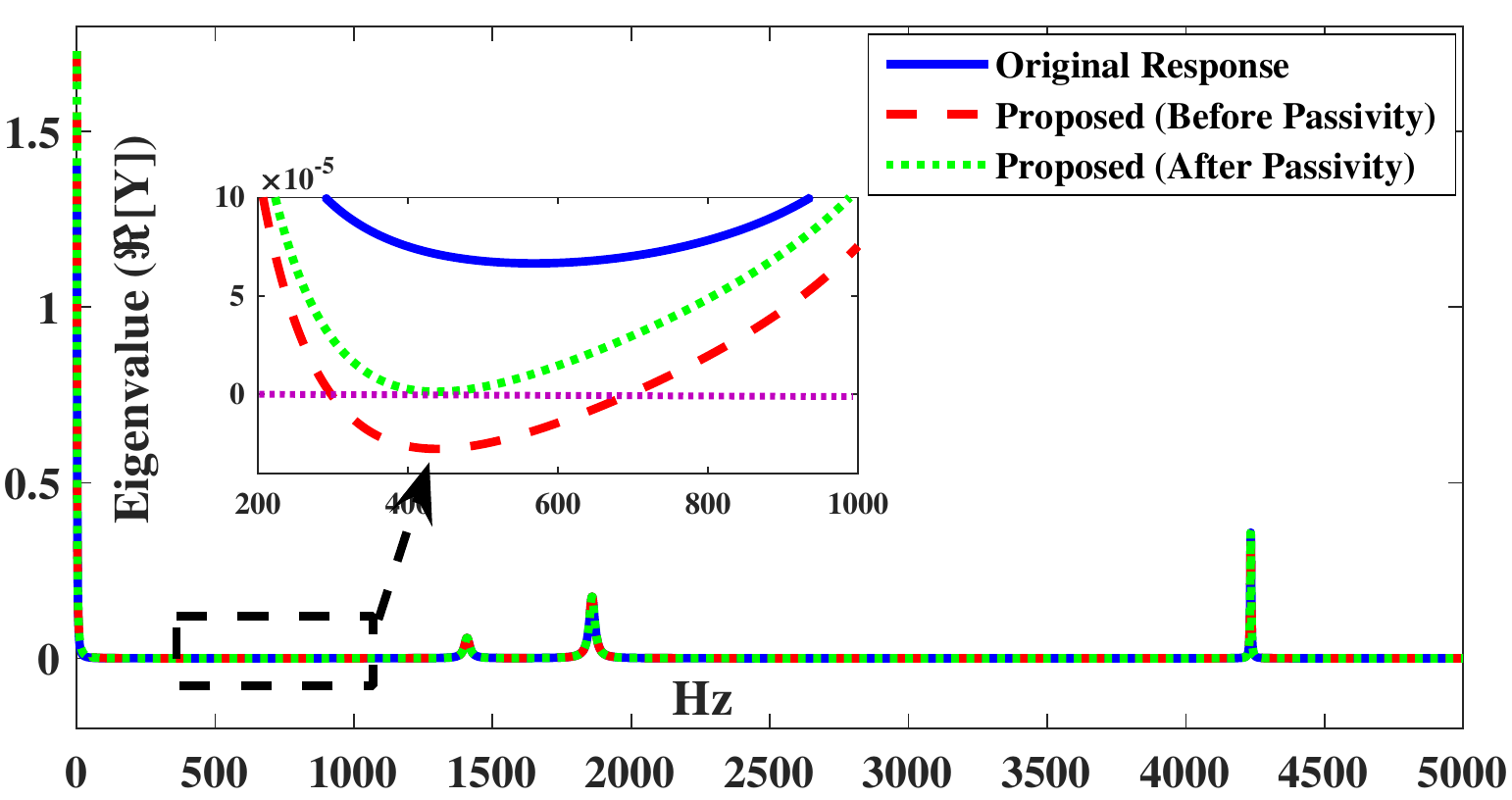}
\caption{The eigenvalue of real-part of admittance matrix (1-port)}
\label{fig13}
\vspace{-4mm}
\end{figure}

Since passivity is related to real part of admittance matrix $(G_{f})$, correcting $G_{f}$  without affecting imaginary part $(B_{f})$ is sufficient. Then $Y_{c}$ can be represented as follows 
\begin{equation}
\label{eqn35}
Y_{c}(z^{-1})=G_{c}(z^{-1})+B_{f}(z^{-1})
\end{equation}
where
\begin{equation}
\label{eqn36}
G_{c}(z^{-1})=G_{f}(z^{-1})+\Delta G
\end{equation}
The objective here is to calculate $\Delta G$. A real, symmetric matrix $G_{b}$  is said to be positive definite if $x^TG_{b}x$ $>$ $0$ $\forall$ $x$ $\ne$ $0$. Thus $x^TG_{b}x$  can be written as 
\begin{equation} \label{eqn37}
\begin{split}
x^TG_{b}x & = \frac{1}{2}x^T\left(G_{b}+G_{b}^T\right)x
\end{split}
\end{equation}
This shows $G_{b}$   is positive definite if and only if $G_{b}+G_{b}^T$ is positive definite. This can be achieved by minimizing an objective function through optimization as
\begin{equation}
\label{eqn38}
\min{} \left\|G_{d}-G_{b}\right\|_F
\end{equation}
\begin{equation}
\label{eqn39}
\textrm{s.t.}\  G_{b}+G_{b}^T > 0
\end{equation}
where $F$  stands for Forbenius norm of a matrix.
We propose a convex optimization formulation to find $G_{b}$ using semi definite programming (SDP) \cite{ref18}. The optimization solution is then used to calculate $\Delta G$ using \eqref{eqn40}. 
\begin{equation}
\label{eqn40}
\Delta G = G_{b}-G_{d}
\end{equation}
The validity of the proposed algorithm is implemented for enforcing passivity of 1-port network formulated previously. From Fig. \ref{fig13} it can be seen that the passivity is enforced. 
\vspace{-4mm}

\subsubsection{Interfacing FDNE with real time simulator}
FDNE can be directly implemented since it is computed in $z$-domain. The implementation process is as follows. With boundary bus voltage $(V_b)$ as input to FDNE, and with $n^{th}$  order estimation, \eqref{eqn7} can be written as
\begin{equation} \label{eqn41}
\begin{split}
I_F(k) & = -a_1I_F(k-1)-a_2I_F(k-2)\dots -a_nI_F(k-n) \\
  & +b_1V_b(k-1)+b_2V_b(k-2)\dots +b_nV_b(k-n)
\end{split}
\end{equation}
where  $I_F$  is current output from FDNE.
For observing high frequency transients only FDNE part is required. To maintain boundary bus parameters at initial steady state, a constant current source is injected into the boundary bus as calculated from \eqref{eqn42} \cite{ref3,ref4}. This can be represented as, 
\begin{equation}
\label{eqn42}
I_b\angle \delta_b=\left(\frac{P_b+jQ_b}{V_b\angle \theta_b}\right)^*
\end{equation}
\begin{equation}
\label{eqn43}
I_{binj}\angle \beta_{inj}=I_b\angle \delta_b-Y_{c}(60Hz)V_b\angle \theta_b 
\end{equation}
where $P_b$   and $Q_b$  are the active and reactive power flow respectively from the boundary bus, $V_b$  and $\theta_b$  are the voltage and angle respectively of the boundary bus. Since admittance at a fundamental component of frequency $(Y_{c}(60Hz))$ is included either in \eqref{eqn5} for \textit{EMT+FDNE+TSA Based Model} or in \eqref{eqn42} for \textit{EMT+FDNE Based Model}, the fundamental frequency component must be eliminated from FDNE. This is performed by subtracting $Y(60Hz)V_b\angle \theta_b$  term in \eqref{eqn43} before injecting boundary bus current to remove fundamental frequency component from FDNE.
\vspace{-4mm}

\begin{figure}[h]
\centering
\includegraphics[width=3.5in]{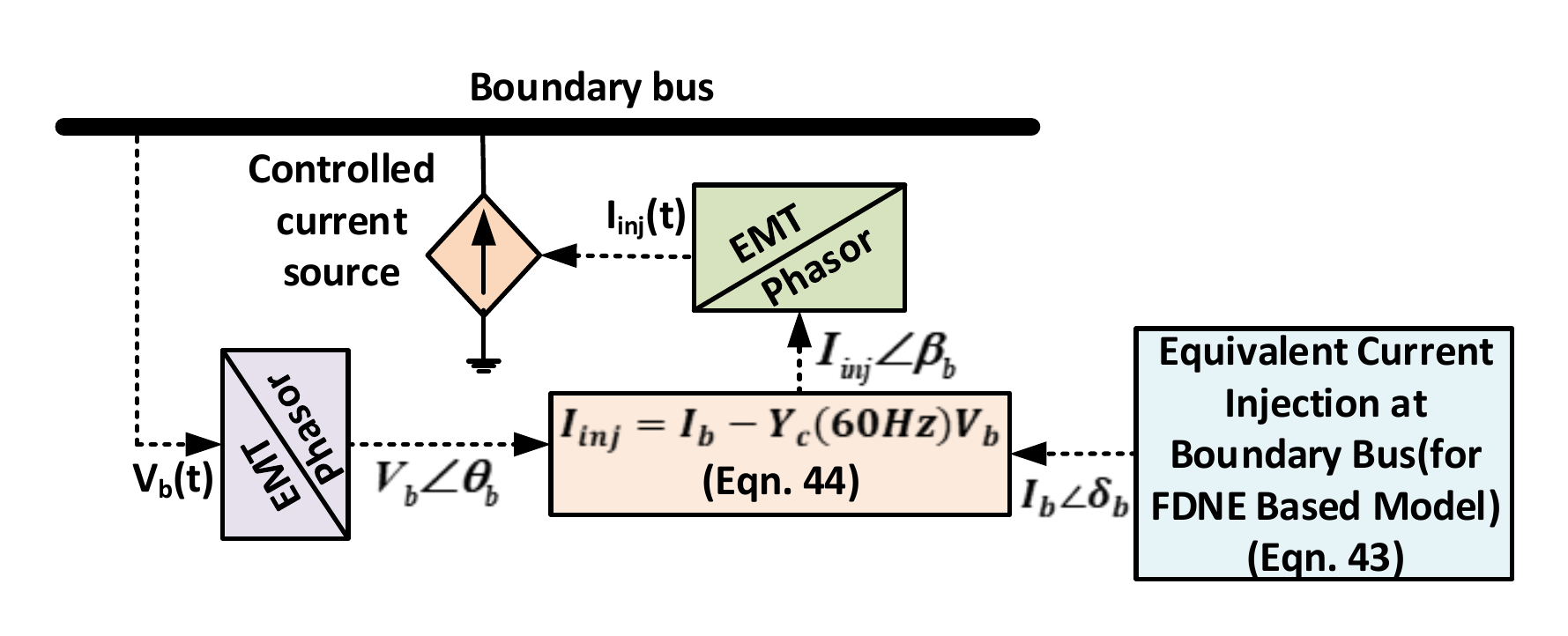}
\caption{Boundary bus current calculation for FDNE only}
\label{fig2y}
\vspace{-5mm}
\end{figure}
\subsection{Study and analysis of FDNE type equivalent modeling}

\begin{figure}[!t]
\centering
\includegraphics[width=3.5in,height=1.4in]{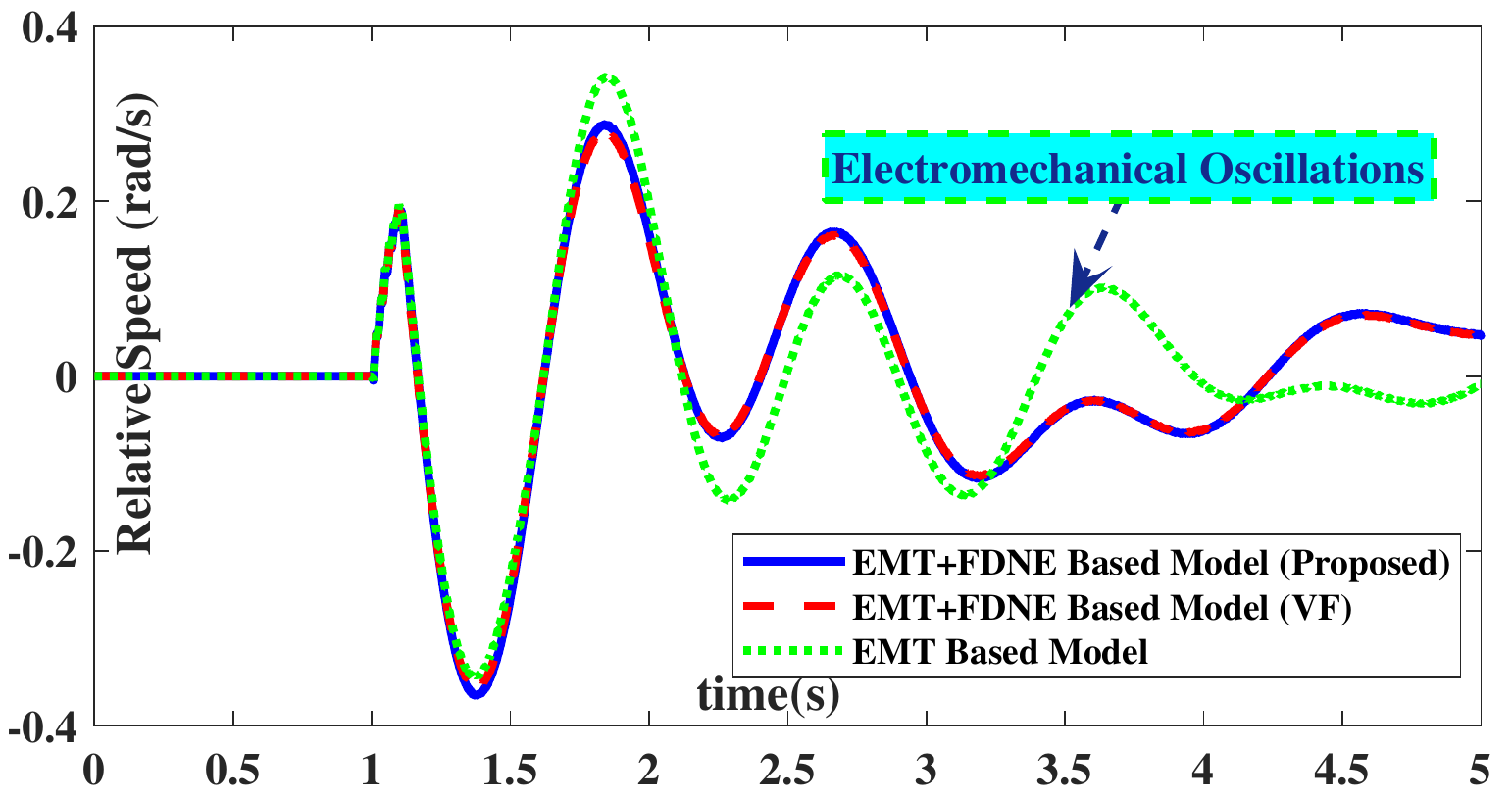}
\caption{Relative speed of Gen.2 w.r.t Gen.1}
\label{fig15}
\vspace{-4mm}
\end{figure}

\begin{figure}[!t]
\centering
\includegraphics[width=3.5in,height=1.4in]{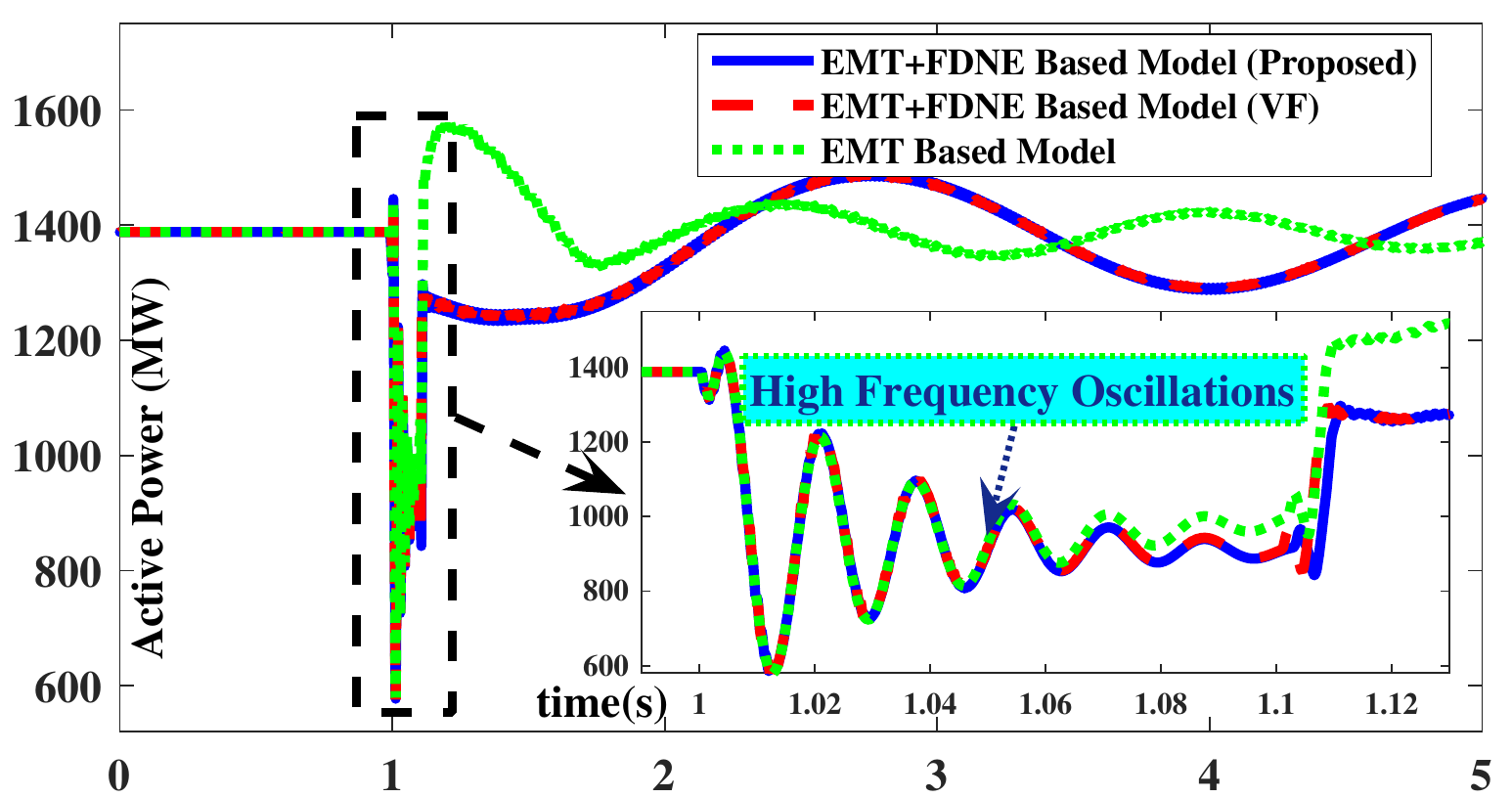}
\caption{Active power flow from bus 10 to bus 9 (Boundary Bus)}
\label{fig16}
\end{figure}

For study and preliminary analysis, the proposed FDNE algorithm is implemented on two area test system. Two cases are considered for study. First case is the proposed method \textit{EMT+FDNE Based Model (Proposed)} and  the  second  case is an existing architecture (\textit{EMT+FDNE Based Model(VF)}).  For  validation,  both the test cases are compared with \textit{EMT Based Model}. For the same type of disturbance as discussed in Section \ref{sec11}, Fig. \ref{fig15} shows the comparison of the relative speed of Gen. 2 w.r.t gen. 1, and Fig. \ref{fig16} shows the comparison of active power flow from bus 10 to bus 9.

\begin{table}[!h]
\renewcommand{\arraystretch}{1.3}
\centering
\caption{Comparison of Reduced (EMT+FDNE) and Original (EMT) Models of Two-Area System}
\label{table3}
\begin{tabular}{*9c}
\hline
 & EMT+FDNE(Proposed) & EMT+FDNE(VF) \\
\hline
Relative Speed (Fig. \ref{fig15}) & 0.481261 & 0.4827015 \\
\hline
Active Power (Fig. \ref{fig16}) & 0.077046 & 0.0771309 \\
\hline
\end{tabular}
\end{table}
\vspace{-2mm}
 Table \ref{table3} shows the comparisons between proposed FDNE and an offline VF based algorithm. Both algorithms gives similar results, proving that FDNE can be formulated online with less computational effort and lower order of transfer function when compared to offline algorithms. From Fig. \ref{fig15} and Fig. \ref{fig16} it can be seen that with this approach, high frequency oscillations are preserved whereas electromechanical oscillations are not preserved, proving the need to have combined FDNE models with TSA equivalents. 
\vspace{-4mm}


\subsection{Implementing TSA and FDNE on two area power system}
\begin{figure}[!h]
\centering
\includegraphics[width=3.5in,height=2.2in]{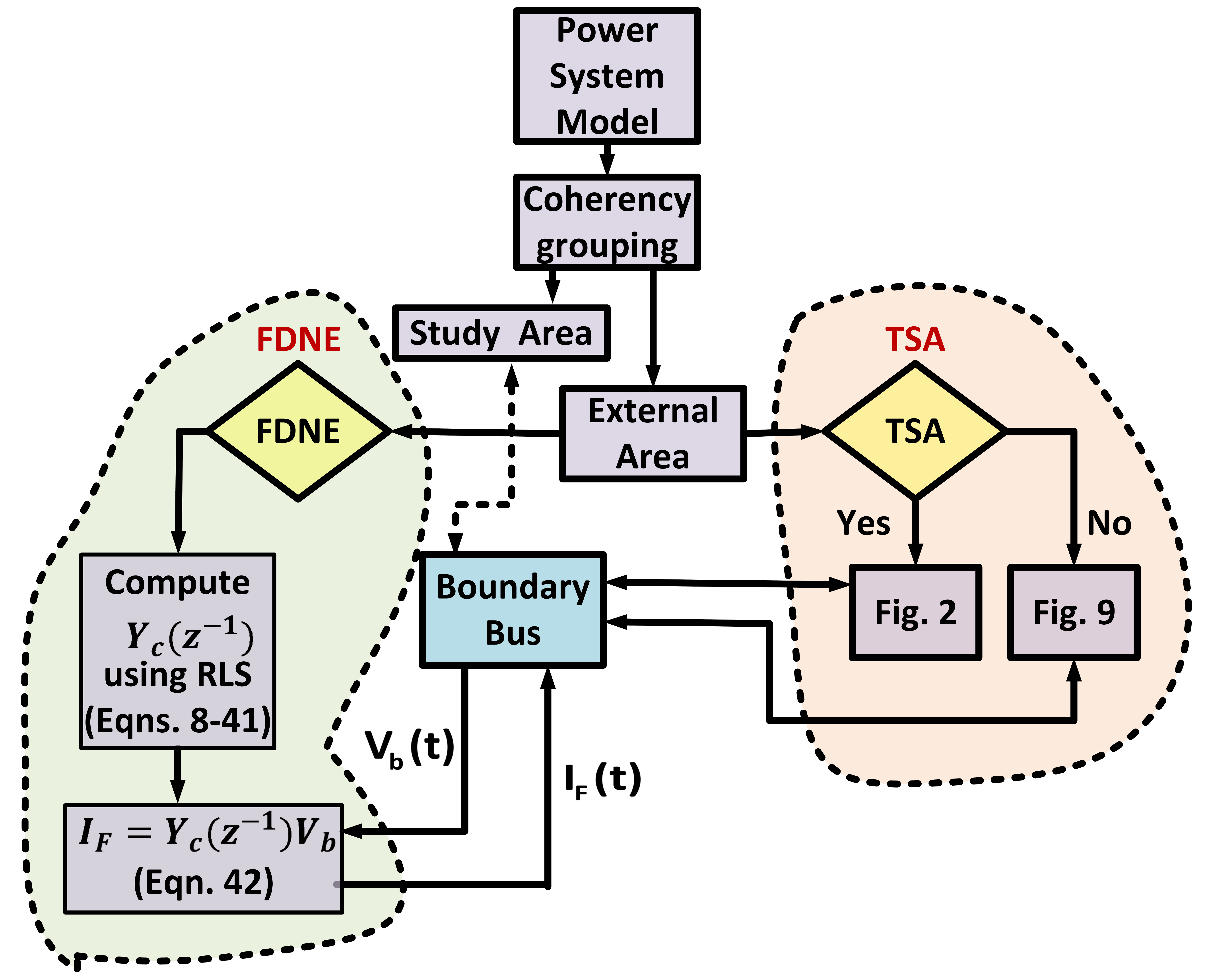}
\caption{Implementation flowchart for TSA and FDNE type models.}
\label{fig17}
\vspace{-3mm}
\end{figure}

\begin{figure}[h]
\centering
\includegraphics[width=3.5in,height=1.4in]{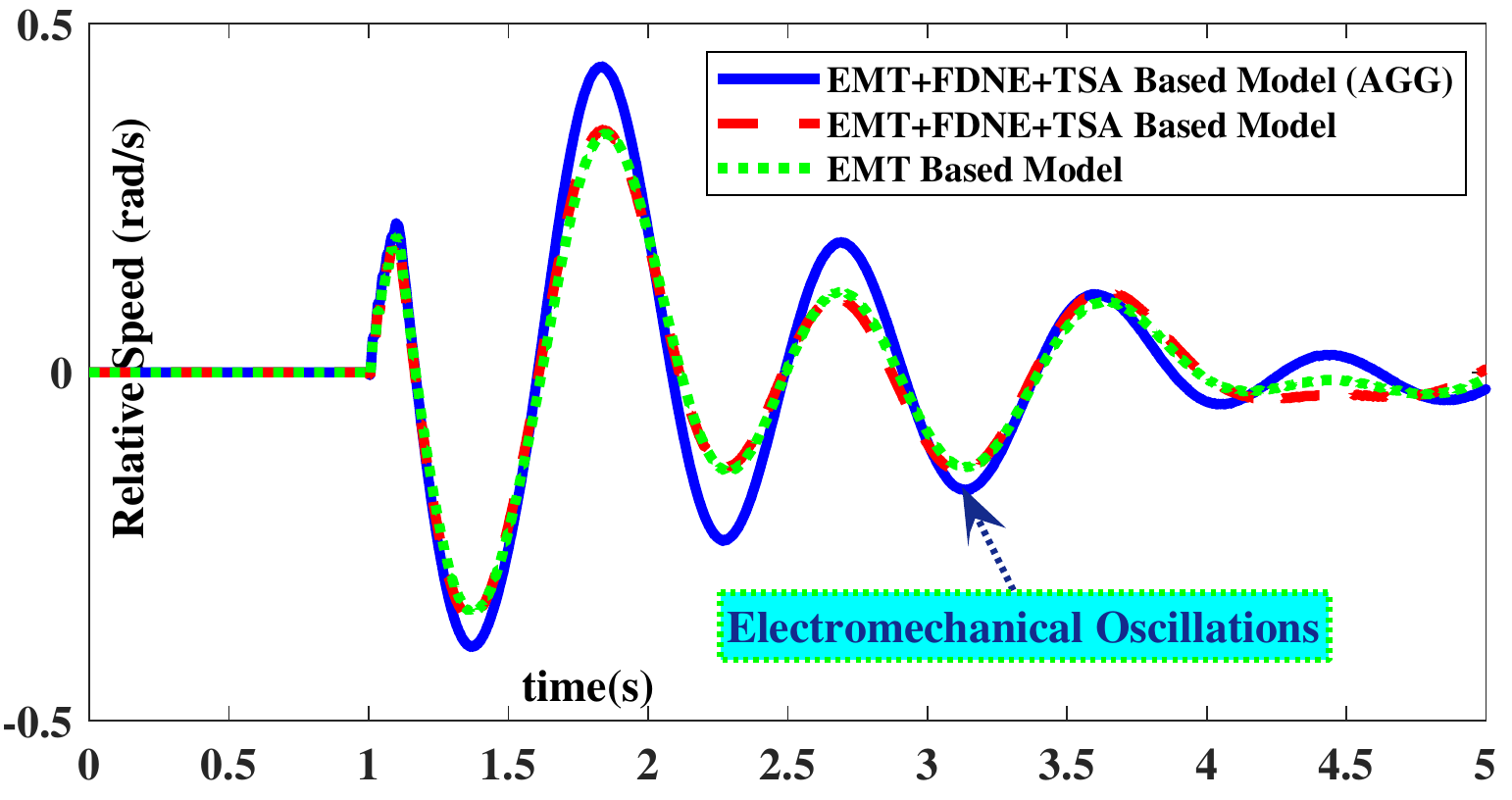}
\caption{Relative speed of Gen.2 w.r.t Gen.1}
\label{fig18}
\vspace{-4mm}
\end{figure}

\begin{figure}[!h]
\centering
\includegraphics[width=3.5in,height=1.4in]{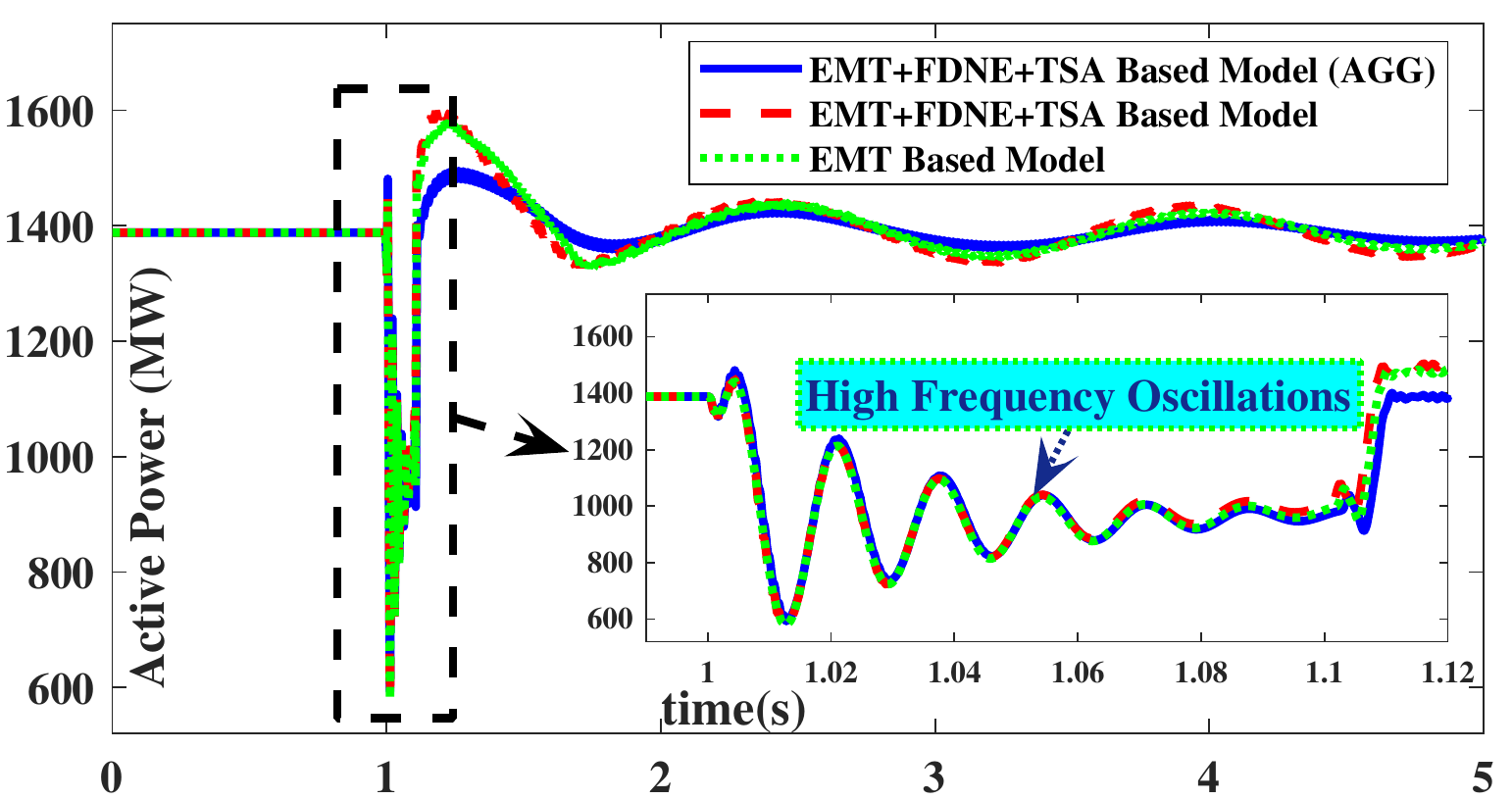}
\caption{Active power flow from bus 10 to bus 9 (Boundary Bus Power)}
\label{fig19}
\vspace{-4mm}
\end{figure}

Fig. \ref{fig17} shows the implementation approach for combined TSA and FDNE type equivalents. Table. \ref{table4} shows the comparison of frequency and damping factor for original model and various reduced order model using eigen value realization algorithm for generator-3 speed data. It can be seen that the proposed approach provides very close result compared to full EMT based model. Fig. \ref{fig18} shows the comparison of the relative speed of Gen. 2 w.r.t gen. 1, and Fig. \ref{fig19} shows the comparison of active power flow from bus 10 to bus 9. From Fig. \ref{fig18} and Fig. \ref{fig19}, it can be seen that both high-frequency and electromechanical oscillations are well preserved. Table. \ref{table3aa} shows the error comparison between orignial and reduced model using \eqref{eqn44}.
\vspace{-4mm}
\vspace{-4mm}
\begin{table}[!h]
\renewcommand{\arraystretch}{1.3}
\centering
\caption{Comparison of Reduced Order (FDNE Only) Models}
\label{table4}
\begin{tabular}{p{2.2cm} c c c}
\hline
Case &  Eigen Value &  Frequency(Hz) & Damping(\%)\\
\hline
\shortstack{EMT Based Model \\ \ } & \shortstack{-1.323$\pm$i7.544 \\ -0.880$\pm$i5.157} & \shortstack{0.6566 \\ 0.8327} & \shortstack{21.92 \\ 16.82}\\
\hline
\shortstack{EMT+TSA Based \\ Model(AGG)} & \shortstack{-0.906$\pm$i7.096 \\ -1.381$\pm$i6.537} & \shortstack{0.5494 \\ 1.0634} & \shortstack{43.09 \\ 20.67}\\
\hline
\shortstack{EMT+TSA Based \\ Model} & \shortstack{-1.306$\pm$i7.583 \\ -0.598$\pm$i5.014} & \shortstack{0.6501 \\ 0.8038} & \shortstack{5.7543 \\ 11.8466}\\
\hline
\shortstack{EMT+FDNE Based \\ Model} & \shortstack{-0.664$\pm$i6.993 \\ -0.373$\pm$i2.507} & \shortstack{0.3822 \\ 0.4034} & \shortstack{61.26 \\ 61.26}\\
\hline
\shortstack{EMT+FDNE+TSA \\Based Model(AGG)} & \shortstack{-1.321$\pm$i7.652 \\ -0.670$\pm$i4.984} & \shortstack{0.6623 \\ 0.8804} & \shortstack{4.1457 \\ 13.336}\\
\hline
\shortstack{EMT+FDNE+TSA \\Based Model} & \shortstack{-1.325$\pm$i7.601 \\ -0.610$\pm$i5.002} & \shortstack{0.6518 \\ 0.8021} & \shortstack{5.2343 \\ 12.1216}\\
\hline
\end{tabular}
\end{table}

\vspace{-5mm}
\begin{table}[!h]
\renewcommand{\arraystretch}{1.3}
\centering
\caption{Comparison of Reduced (EMT+FDNE+TSA) and Original (EMT) Models of Two-Area System}
\label{table3aa}
\begin{tabular}{*9c}
\hline
 & \shortstack{EMT+FDNE+TSA \\ (AGG)} & \shortstack{EMT+FDNE+TSA} \\
\hline
Relative Speed (Fig. \ref{fig18}) & 0.2286 & 0.0923\\
\hline
Active Power (Fig. \ref{fig19}) & 0.0163 & 0.0076 \\
\hline
\end{tabular}
\end{table}


\section{Implementation Test on Interconnected Power Grid}
 To prove scalability and implementation using a multi-port networks,  IEEE 39 and 68 bus power system models are considered. Plesae refer \cite{ref20} and \cite{ref5a} for IEEE 39 and 68 bus system details and one-line diagrams respectively.
 
 \subsection{IEEE 39 Bus System Implementation Test Results}
\begin{figure}[!h]
\centering     
\subfigure[Magnitude of admittance ($Y_{33}$)]{\label{fig11}\includegraphics[width=1.72in,height=1.4in]{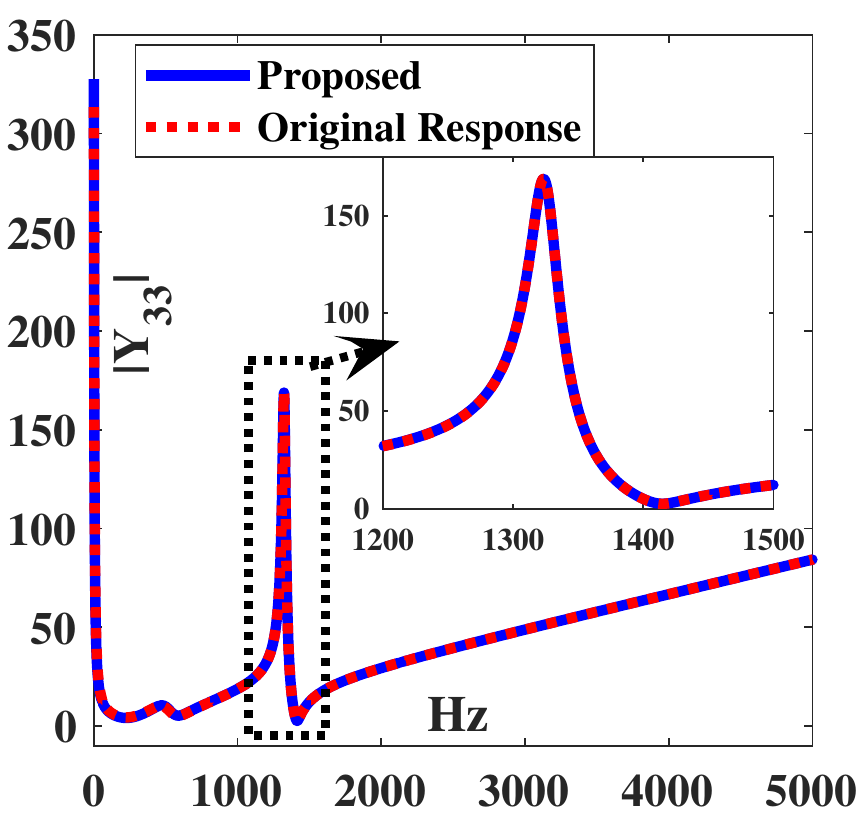}}
\subfigure[Angle of admittance ($Y_{33}$)]{\label{fig12}\includegraphics[width=1.72in,height=1.4in]{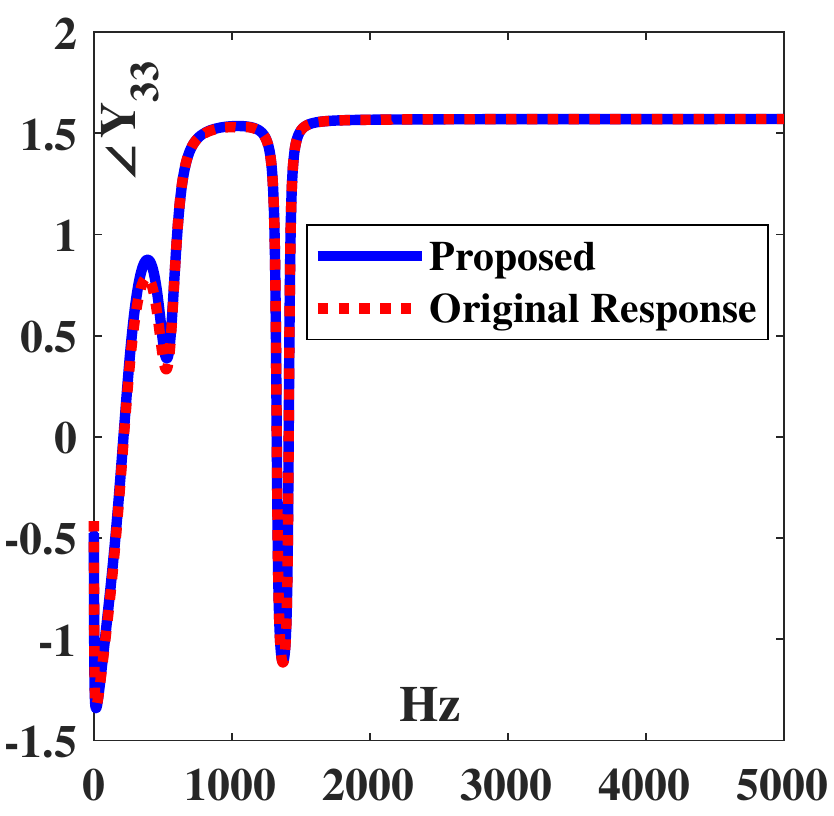}}
\caption{Admittance vs frequency of external area (39 Bus System)}
\vspace{-2mm}
\end{figure}

\begin{figure}[!h]
\centering
\includegraphics[width=3.5in,height=1.6in]{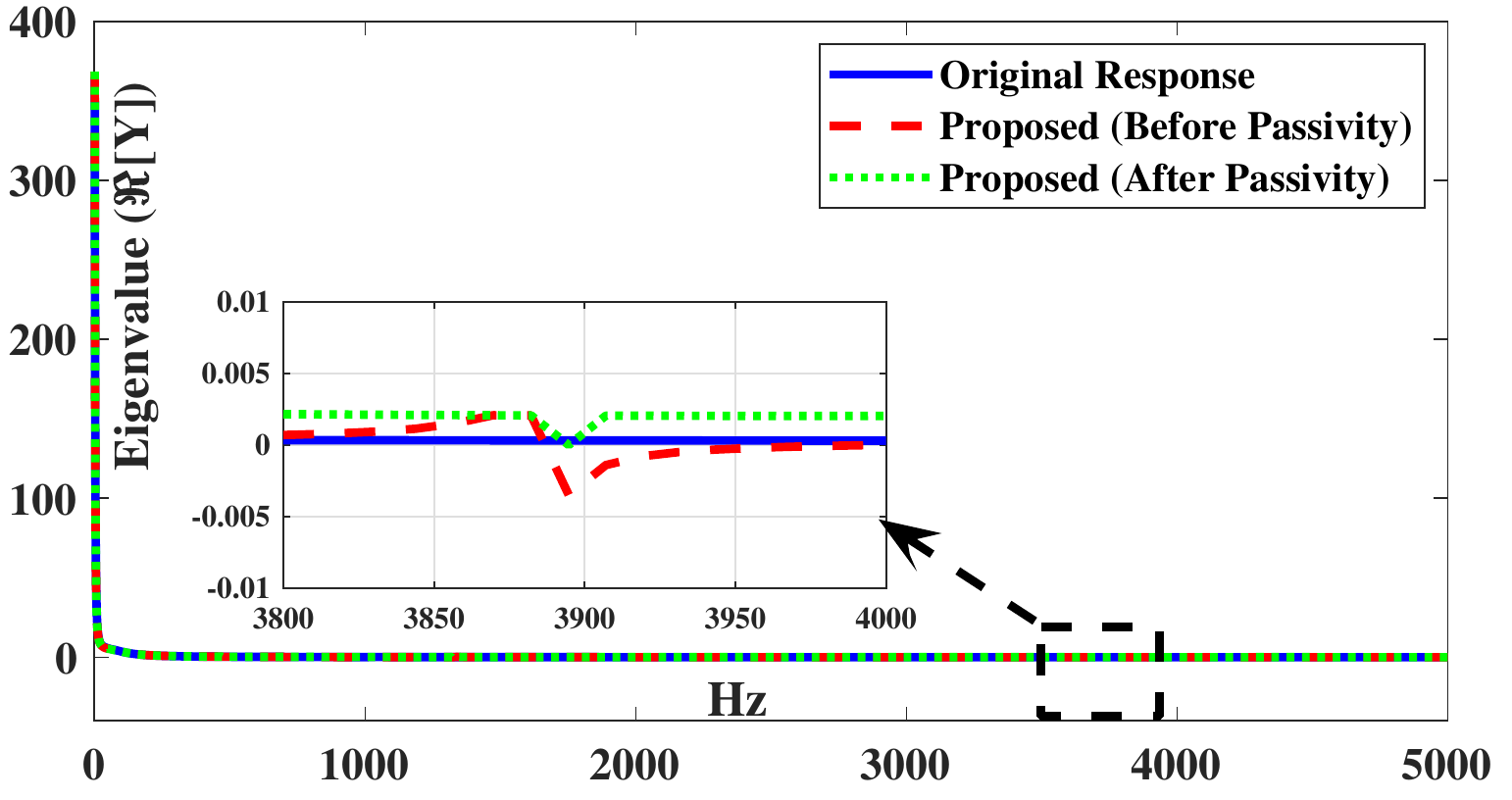}
\caption{The corrected eigenvalue of real-part of admittance matrix (3-port)}
\label{fig14}
\vspace{-4mm}
\end{figure}

 In this system first, based on the coherency grouping of the generators the test system is divided into \textit{study} and \textit{external} area. To assess the performance of proposed \textit{EMT+FDNE+TSA} based reduced order model, Group-I which consists of generators 4, 5, 6, 7, 9 is considered as \textit{external} area and the rest of the power system as \textit{study} area. The \textit{study} and \textit{external} area is divided at bus 16 (Port-1), bus 17 (Port-2), and bus 26 (Port-3). Fig. \ref{fig11} and Fig. \ref{fig12} shows the admittance magnitude and angle of port-3. Fig. \ref{fig14} shows the passivity enforcement for 3-port network.
 \begin{figure}[!h]
\centering
\includegraphics[width=3.5in,height=1.35in]{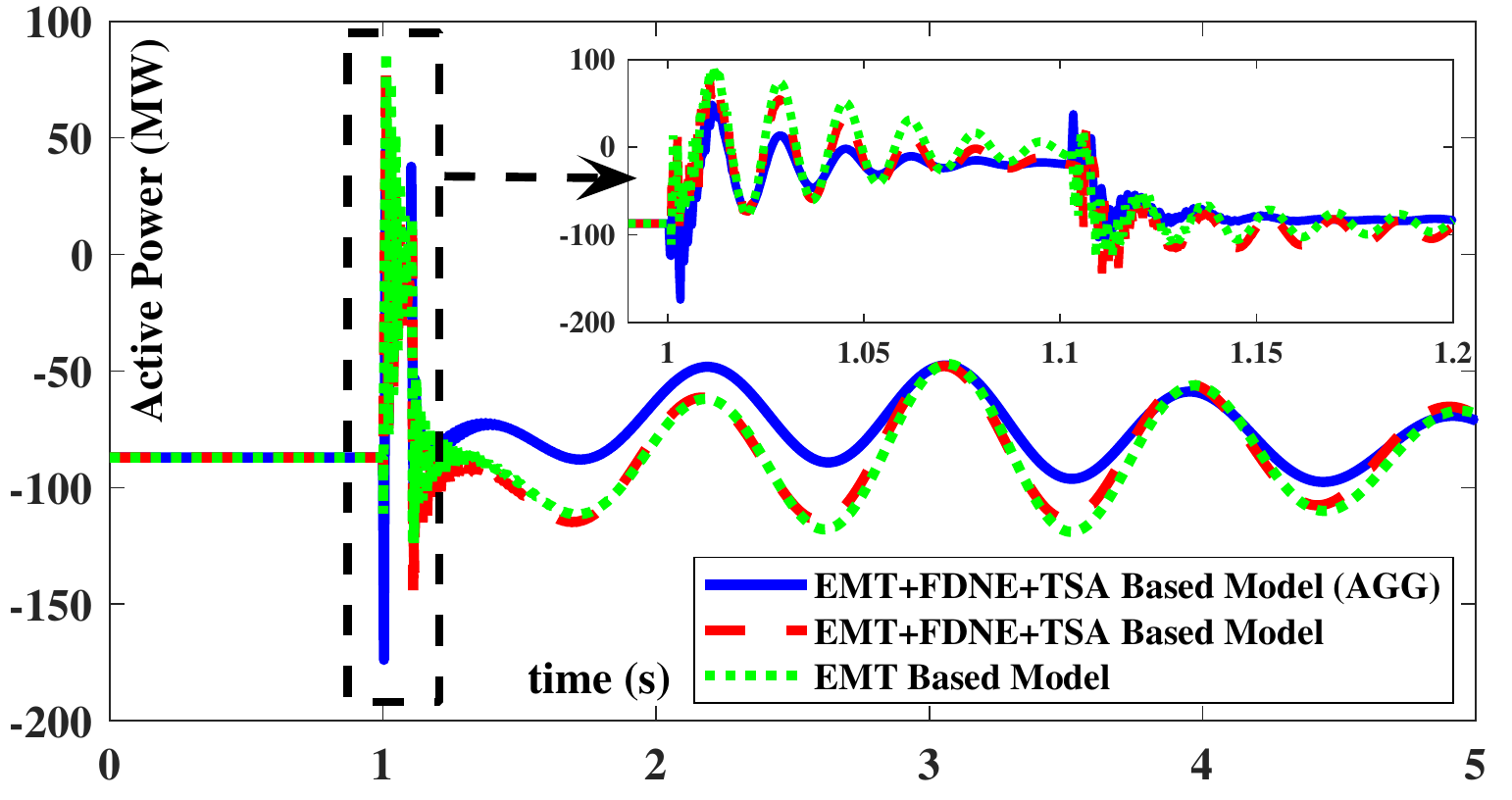}
\caption{Bus 26 Active Power}
\label{fig20}
\vspace{-4mm}
\end{figure}
 \begin{figure}[!h]
\centering
\includegraphics[width=3.5in,height=1.35in]{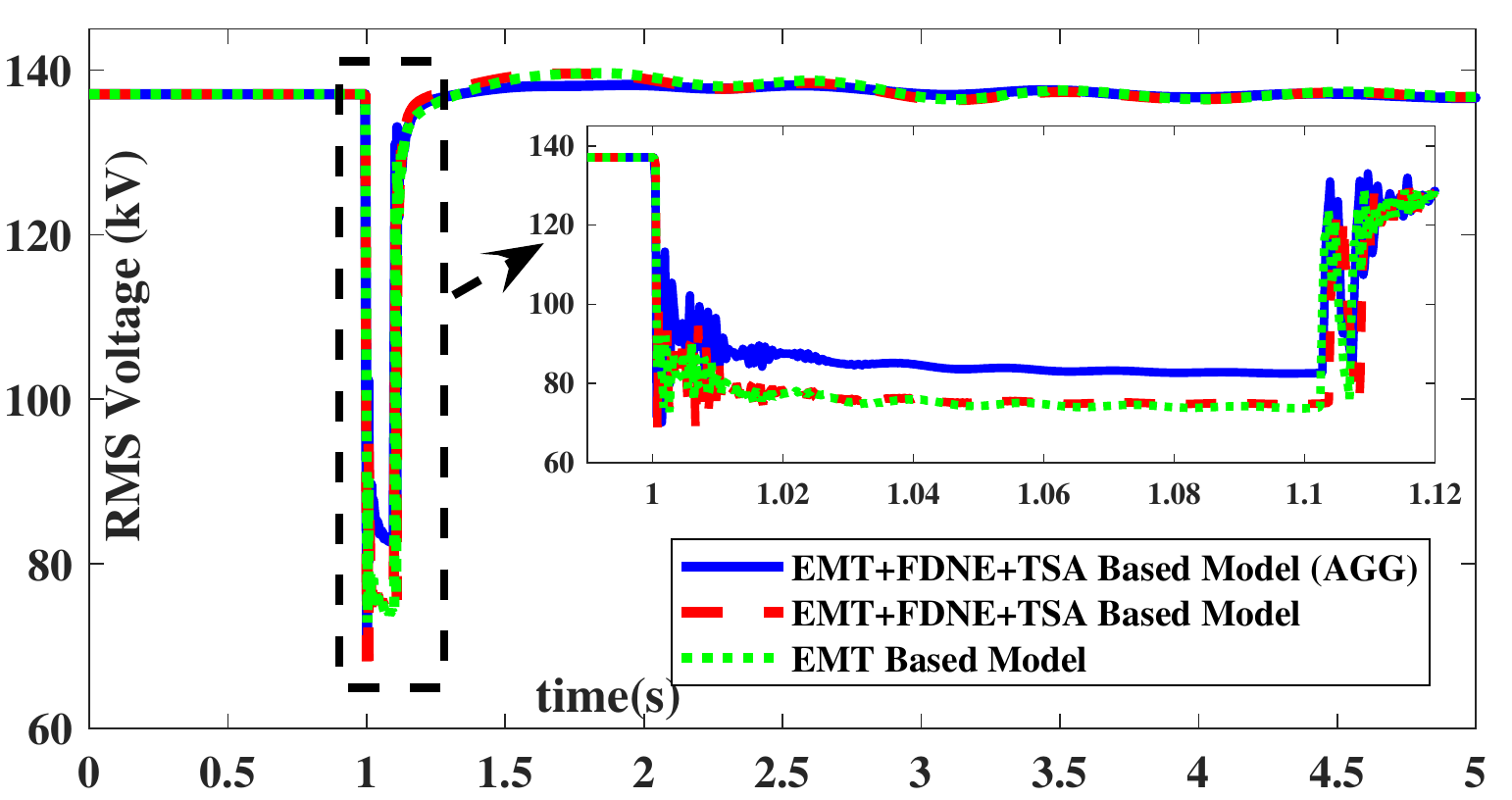}
\caption{Bus 17 Voltage}
\label{fig22}
\vspace{-4mm}
\end{figure}
\begin{figure}[!h]
\centering
\includegraphics[width=3.5in,height=1.4in]{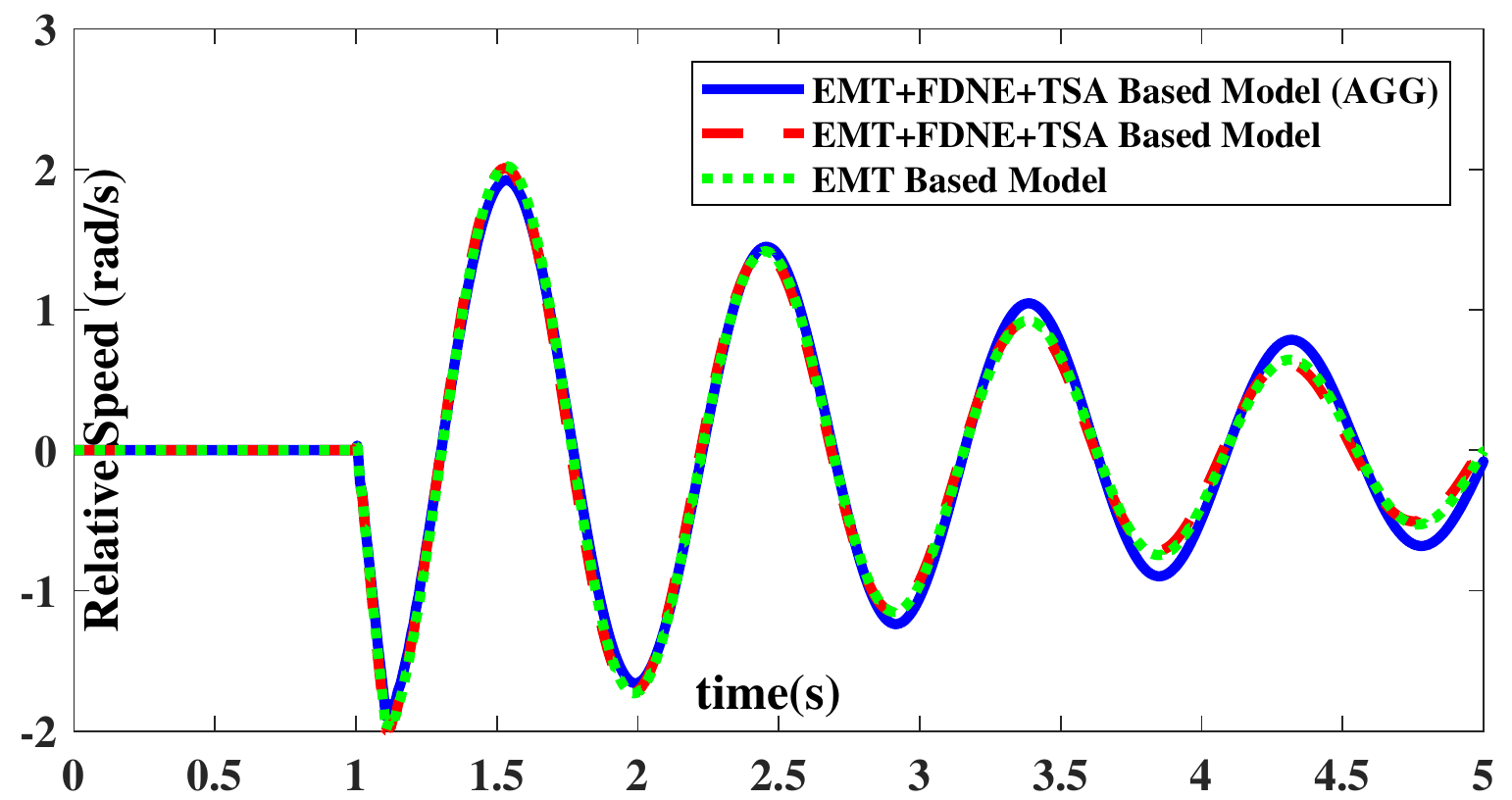}
\caption{Relative speed of generator gen.1 w.r.t gen.2}
\label{fig23}
\vspace{-4mm}
\end{figure}

\begin{table}[!h]
\renewcommand{\arraystretch}{1.3}
\centering
\caption{Comparison of Reduced (EMT+FDNE+TSA) and Original (EMT) Models of IEEE 39 Bus System}
\label{table6}
\begin{tabular}{*9c}
\hline
 & \shortstack{EMT+FDNE+TSA \\ (AGG)} & \shortstack{EMT+FDNE+TSA} \\
\hline
Active Power (Fig. \ref{fig20}) & 0.1634 & 0.0489\\
\hline
Voltage (Fig. \ref{fig22}) & 0.0112 & 0.0081 \\
\hline
Relative Speed (Fig. \ref{fig23}) & 0.0905 & 0.0433 \\
\hline
\end{tabular}
\end{table}

For analysis, a three-phase fault is initialized for a duration of 0.1s and the simulation results are compared with the original model. Fig. \ref{fig20} to Fig. \ref{fig23} shows the validation results of the proposed algorithm. Table \ref{table6} shows relative error \eqref{eqn44} comparison of reduced and original models. It can be seen that the reduced model closely resembles the full EMT model.
\vspace{-4mm}
\subsection{IEEE 68 Bus System Implementation Test Results}
\begin{figure}[!h]
\centering     
\subfigure[Magnitude of admittance ($Y_{22}$)]{\label{fig11x}\includegraphics[width=1.72in,height=1.35in]{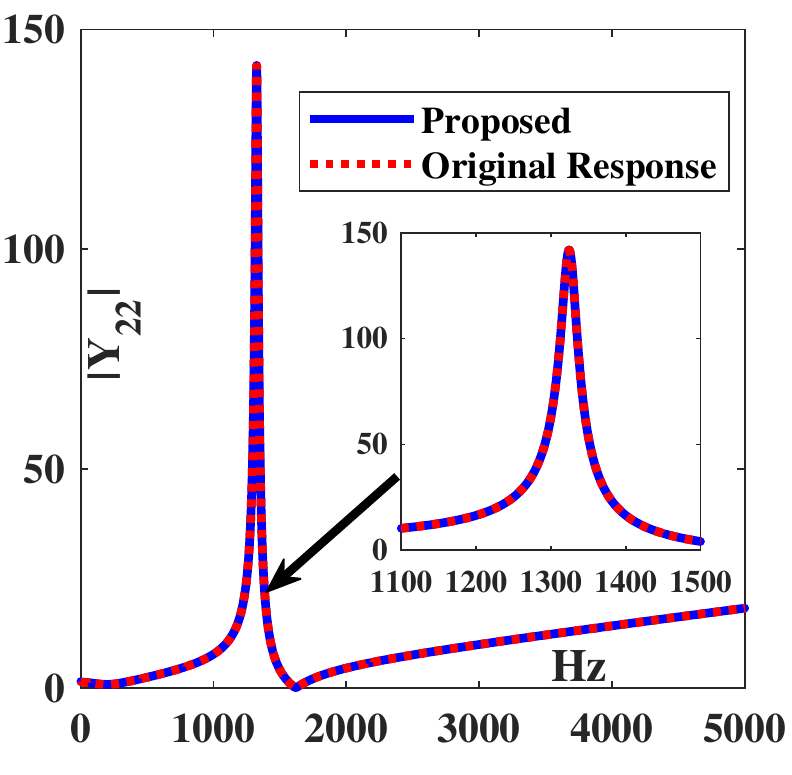}}
\subfigure[Angle of admittance ($Y_{22}$)]{\label{fig12x}\includegraphics[width=1.72in,height=1.30in]{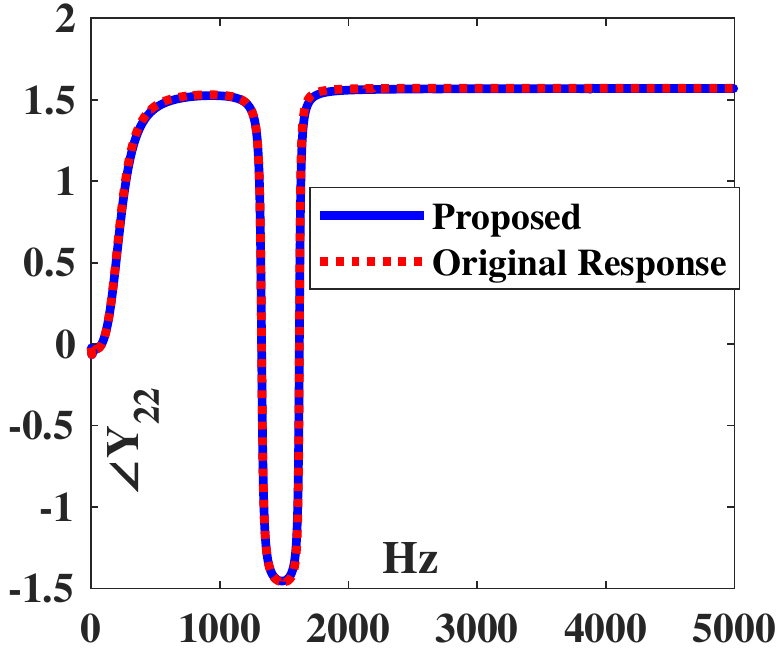}}
\caption{Admittance vs frequency of external area (68 Bus System)}
\vspace{-2mm}
\end{figure}

\begin{figure}[!h]
\centering
\includegraphics[width=3.5in,height=1.3in]{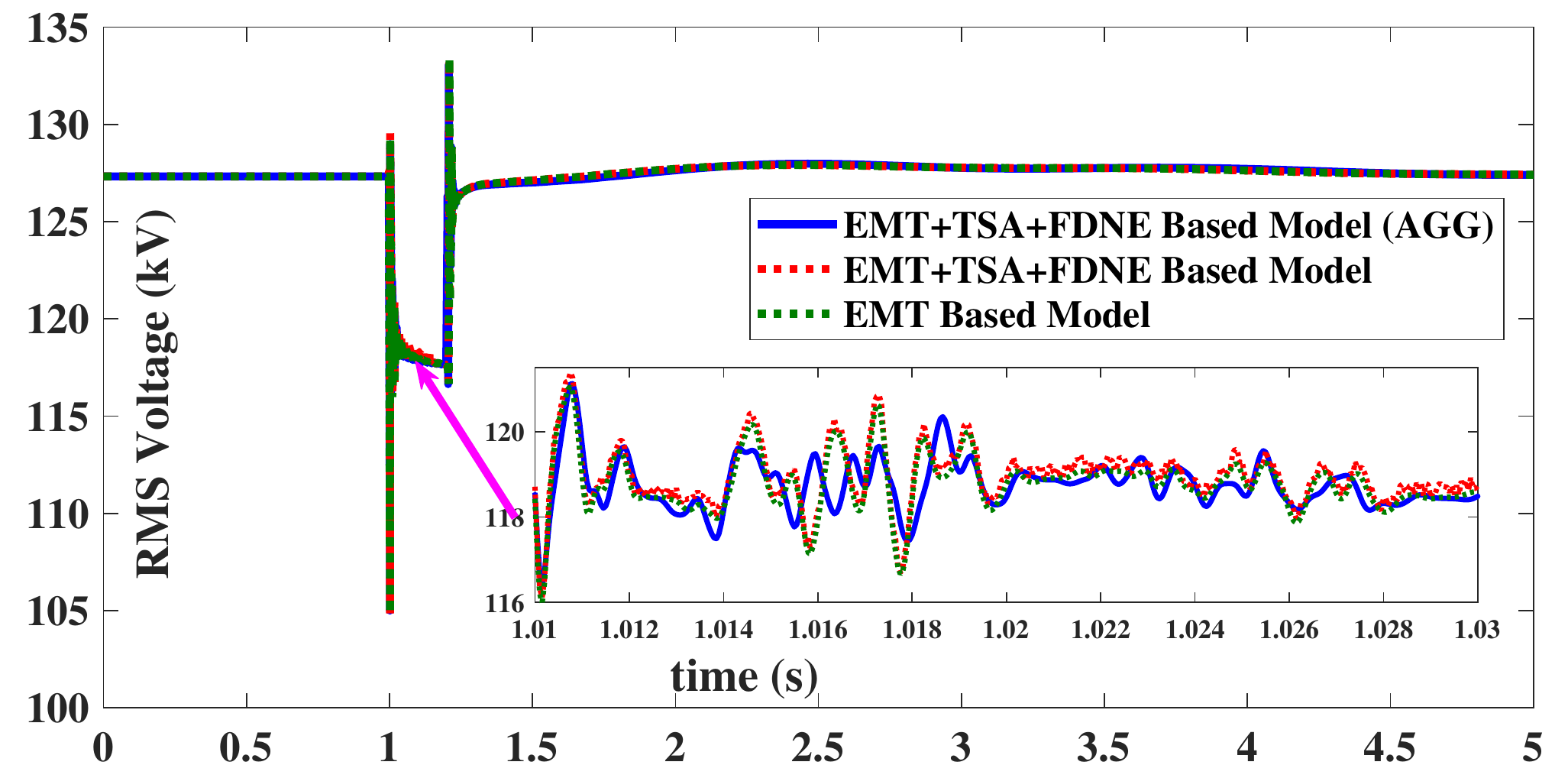}
\caption{Bus 60 Voltage}
\label{fig22x}
\vspace{-4mm}
\end{figure}

\begin{figure}[!h]
\centering
\includegraphics[width=3.5in,height=1.3in]{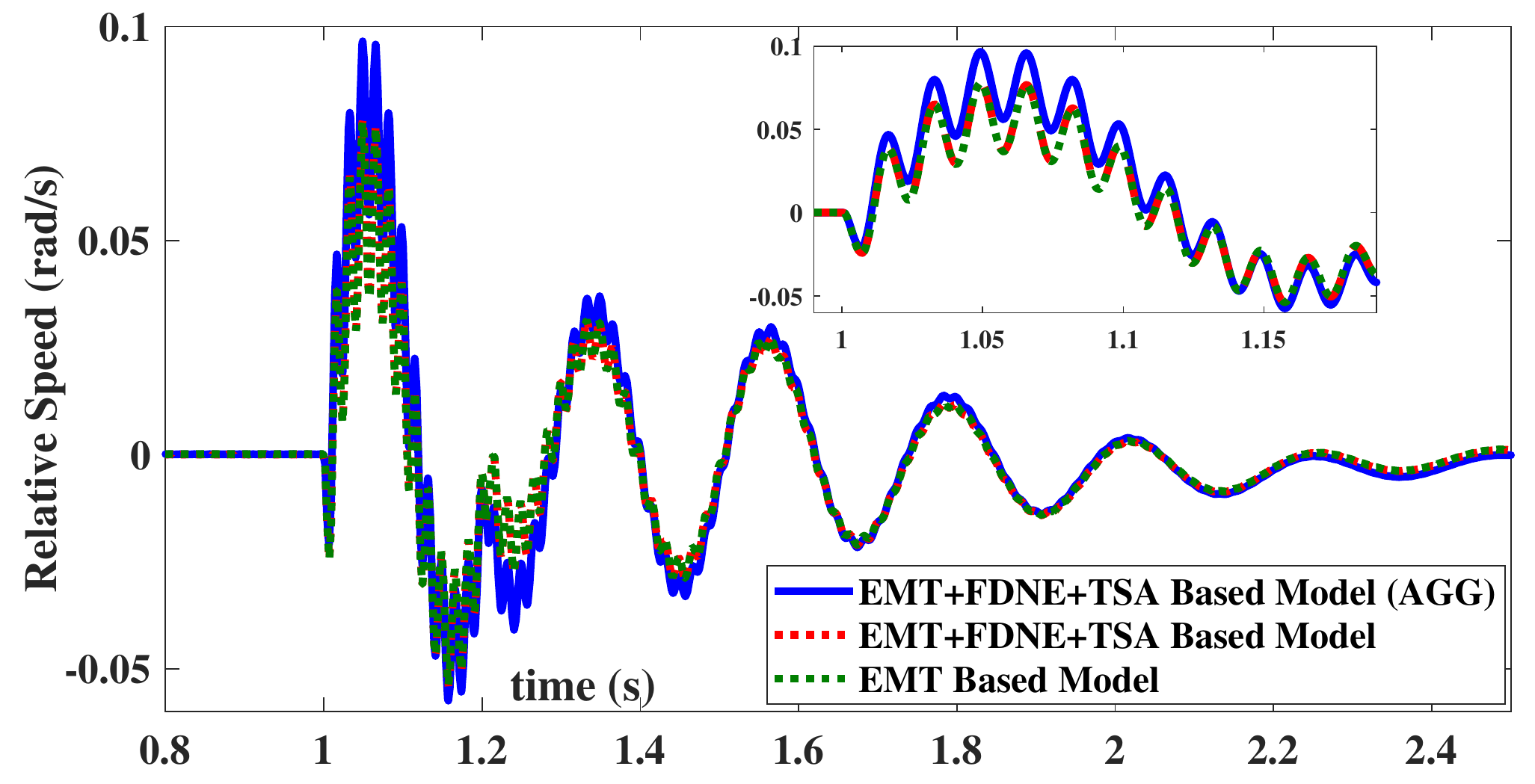}
\caption{Relative speed of generator 10 w.r.t generator 16}
\label{fig23x}
\vspace{-4mm}
\end{figure}

Further, in order to validate the proposed algorithm on a larger system, IEEE 68 bus system which consists of 16 generators consists of 5-areas is considered. The area with generators 1 to 9 is considered as $external$ area. The $study$ and $external$ areas are divided at bus 54 (Port-2), 27 (Port-3), and 60 (Port-1). Fig. \ref{fig11x} and Fig. \ref{fig12x} shows the admittance magnitude and angle of port-2. For analysis a 3-ph fault for a duration of 0.2s is simulated at bus-49 starting at at 1s. The responses of reduced order model are compared with the original model. Fig. \ref{fig22x} shows the Bus-60 voltage and Fig. \ref{fig23x} shows the relative speed of generator-10 w.r.t generator-16. 

From the above comparisons it can be seen that the dynamics of the proposed reduced order model is similar to that of the original model. The computational time required to generate equivalents is shown in Table. \ref{table4aa}.

\begin{table}[!h]
\renewcommand{\arraystretch}{1.3}
\centering
\caption{Time Required for Generation of Equivalents}
\label{table4aa}
\begin{tabular}{*9c}
\hline
 Model & TSA(AGG) & TSA & FDNE \\
\hline
Two-Area System & 10s & 8s & 40s\\
\hline
39 Bus System & 14s & 13s & 91s\\
\hline
68 Bus System & 19s & 17s & 97s\\
\hline
\end{tabular}
\end{table}
\vspace{-7mm}

\section{Conclusion}
In this paper a novel real-time frequency based reduced order modeling of the large power system for EMT simulation is proposed. In the proposed architecture, external area is modeled as a combination of FDNE and TSA. FDNE is formulated using an online discrete RLS which can preserve high-frequency behavior, where as TSA preservers the electromechanical (low frequency) behavior of the system under consideration. The approach also enforces passivity conditions. Implementation results in two area, IEEE 39 and 68 bus power system models shows that the proposed reduced order model represents the aggregated power grid accurately and at the same time can capture the oscillations as close as the full detail model for EMT simulations. 
\vspace{-4mm}

%

\ifCLASSOPTIONcaptionsoff
  \newpage
\fi



%
\bibliographystyle{IEEEtran}
\bibliography{Ref.bbl}
\input{Ref.bbl}

\vspace{-8 mm}
\begin{IEEEbiography}[{\includegraphics[width=1in,height=1.25in,clip,keepaspectratio]{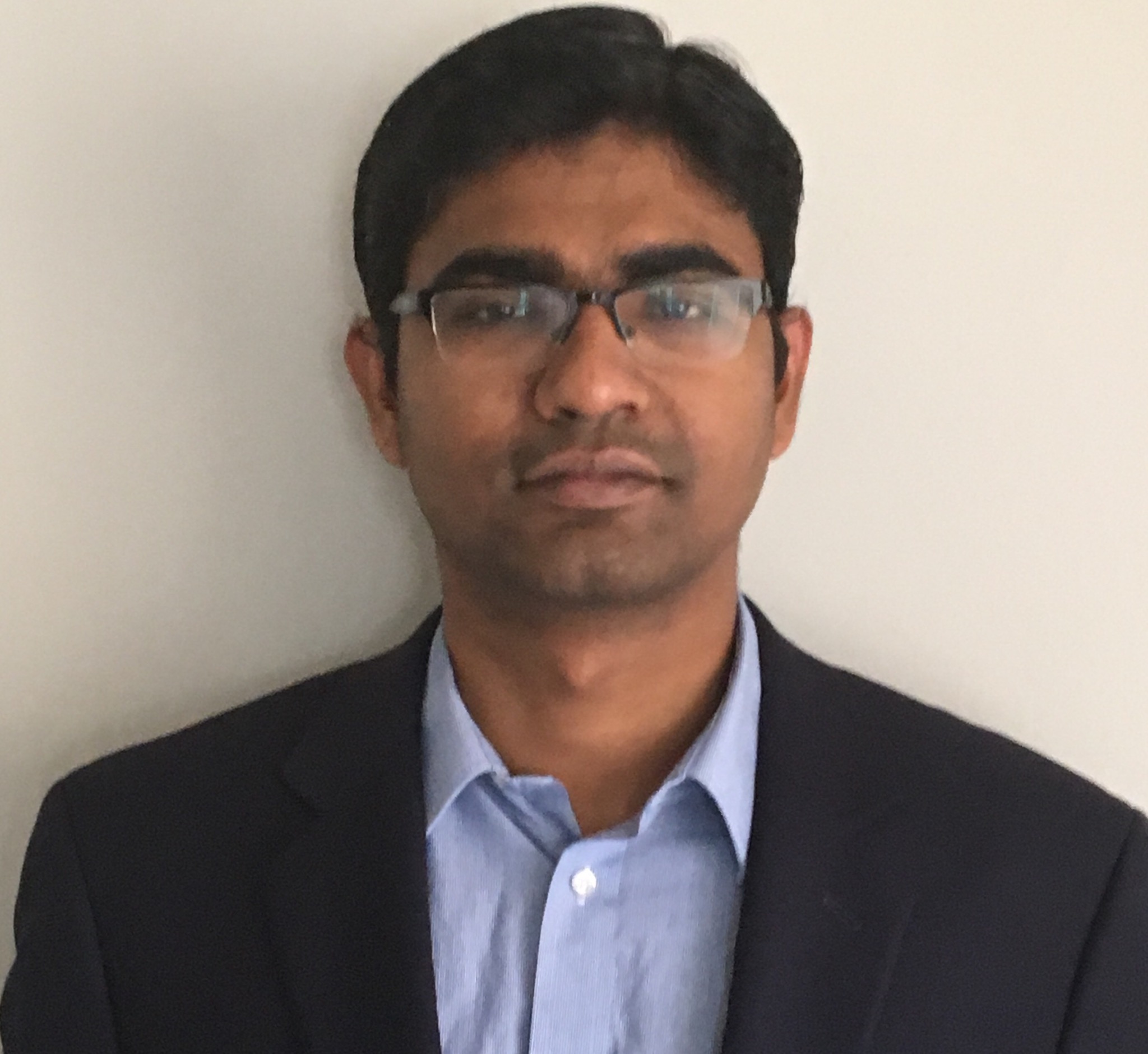}}]{A. Thakallapelli}(S'14) received his B.Tech degree in Electrical Engineering from Acharya Nagarjuna University in 2010 and the M.Tech degree in Electrical Engineering from the Veermata Jijabai Technological Institute in 2012. He is currently working toward the Ph.D degree in Electrical Engineering from the Department of Electrical and Computer Engineering, University of North Carolina at Charlotte. His research interests include wide-area control, reduced order modeling, power system stability and renewable energy.
\end{IEEEbiography}
\vspace{-13 mm}
\begin{IEEEbiography}[{\includegraphics[width=1in,height=1.25in,clip,keepaspectratio]{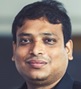}}]{S. Ghosh}(M'12) received the B.SC Engg.from National Institute of Technology (NIT) Jamshedpur, India, in 2002, the M.Tech degree from NIT Durgapur, India, in 2009, and the Ph.D. degree from Indian Institute of Technology, Delhi, India, in 2013. Currently, he is working in Khalifa University, Abu Dhabi. He worked as an Associated Graduate Faculty member with the Department of Electrical and Computer Engineering, University of North Carolina, Charlotte, NC, USA and as an  Assistant  Professor  with  the  Department  of  Electrical  Engineering, Indian School of Mines, Dhanbad, India.  His research interests include power system stability, model order reduction, and grid integration with renewable energy resources.
\end{IEEEbiography}
\vspace{-13 mm}
\begin{IEEEbiography}
[{\includegraphics[width=1in,height=1.25in,clip,keepaspectratio]{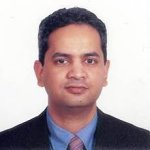}}]{S. Kamalasadan}(S'01, M'05, SM'17) received his B Tech. degree in Electrical and Electronics from the University of Calicut, Kerala, India in 1991, M.Eng in Electrical Power Systems Management, from the Asian Institute of Technology, Bangkok, Thailand in 1999, and Ph.D. in Electrical Engineering from the University of Toledo, Ohio, USA in 2004. He is currently working as a Professor in the department of electrical and computer engineering at the University of North Carolina at Charlotte. He has won several awards including the NSF CAREER award and IEEE best paper award. His research interests include Intelligent and Autonomous Control, Power Systems dynamics, Stability and Control, Smart Grid, Micro-Grid and Real-time Optimization and Control of Power System. 
\end{IEEEbiography}

%








\end{document}